\documentclass[preprint,12pt]{elsarticle}

\usepackage{hyperref}
\usepackage{graphicx}
\usepackage{subcaption}

\usepackage{amssymb}
\usepackage{amsmath}

\usepackage{multirow}

\usepackage[utf8]{inputenc}

\newcommand{\ten}[1]{\ensuremath{\mathbf{#1}}}

\usepackage{lineno}

\journal{}

\begin{document}

\begin{frontmatter}



\title{Entropically Damped Artificial Compressibility for SPH}

\author[IITB]{Prabhu Ramachandran\corref{cor1}}
\ead{prabhu@aero.iitb.ac.in}
\address[IITB]{Department of Aerospace Engineering, Indian Institute of
  Technology Bombay, Powai, Mumbai 400076}

\author[NUMECA]{Kunal Puri} \ead{kunal.r.puri@gmail.com}
\address[NUMECA]{NUMECA International S.A,  Chaussée de la Hulpe, 189, Terhulpsesteenweg, 1170 Brussels, Belgium}

\cortext[cor1]{Corresponding author}

\begin{abstract}

  In this paper, the Entropically Damped Artificial Compressibility (EDAC)
  formulation of Clausen (2013) is used in the context of the Smoothed
  Particle Hydrodynamics (SPH) method for the simulation of incompressible
  fluids. Traditionally, weakly-compressible SPH (WCSPH) formulations have
  employed artificial compressiblity to simulate incompressible fluids. EDAC
  is an alternative to the artificial compressiblity scheme wherein a pressure
  evolution equation is solved in lieu of coupling the fluid density to the
  pressure by an equation of state. The method is explicit and is easy to
  incorporate into existing SPH solvers using the WCSPH formulation. This is
  demonstrated by coupling the EDAC scheme with the recently proposed
  Transport Velocity Formulation (TVF) of Adami et al.~(2013). The method
  works for both internal flows and for flows with a free surface. Several
  benchmark problems are considered to evaluate the proposed scheme and it is
  found that the EDAC scheme gives results that are as good or sometimes
  better than those produced by the TVF or standard WCSPH. The scheme is
  robust and produces smooth pressure distributions and does not require the
  use of an artificial viscosity in the momentum equation although using some
  artificial viscosity is beneficial.

\end{abstract}


\end{frontmatter}




\section{Introduction}
\label{sec:intro}

The Smoothed Particle Hydrodynamics (SPH) technique was initially developed
for astrophysical problems independently by Lucy~\cite{lucy77}, and Gingold
and Monaghan~\cite{monaghan-gingold-stars-mnras-77}. The method is mesh-free
and self-adaptive. With the introduction of the weakly-compressible SPH scheme
(WCSPH) by Monaghan~\cite{sph:fsf:monaghan-jcp94}, the SPH method has been
extensively applied to incompressible fluid flow and free-surface problems
(see~\cite{Shadloo16} and~\cite{Violeau16} for a recent review with an
emphasis on the application of SPH to industrial fluid flow problems).
Alternative to the WCSPH approach, truly incompressible implicit SPH schemes
like the projection-SPH~\cite{sph:psph:cummins-rudman:jcp:1999} and
incompressible-SPH~\cite{isph:shao:lo:awr:2003,isph:hu-adams:jcp:2007} have
been introduced. These methods satisfy the incompressiblity constraint
($\nabla \cdot \ten{u} = 0$) by solving a pressure-Poisson equation. The
methods differ in how the pressure-Poisson equation is setup. While these
schemes are generally considered to be more accurate, the implicit nature of
these schemes makes it difficult to implement and parallelize which has lead
to the WCSPH approach garnering favor within the SPH community. The Implicit
Incompressible SPH scheme (IISPH)~\cite{iisph:ihmsen:tvcg-2014} proposes an
iterative solution procedure to alleviate some of these issues. Recently, an
Artificial Compressibility-based Incompressible SPH (ACISPH)
scheme~\cite{sph:acisph:cpc:2017} has been proposed which applies a
traditional dual-time stepping approach used in Eulerian schemes to satisfy
incompressibility. This scheme is also explicit and seems a promising
alternative to the traditional ISPH schemes.

The weakly-compressible formulation relies on a stiff equation of state that
generates large pressure changes for small density variations. A consequence
is that the large pressure oscillations need to be damped out, which
necessitate the use of some form of artificial viscosity. Another problem with
the WCSPH formulation is the appearance of void regions and particle clumping,
especially where the pressure is negative. This has resulted in some
researchers using problem-specific background pressure values to mitigate this
problem. The Transport Velocity Formulation (TVF) of \citet{Adami2013}
ameliorates some of the above issues by ensuring a more homogeneous
distribution of particles by introducing a background pressure field. This
background pressure is not tuned to any particular problem. In addition, the
particles are moved using an advection (transport) velocity instead of the
actual velocity. The advection velocity differs from the momentum velocity
through the addition of the constant background pressure. The motion induced
by the background pressure is corrected by introducing an additional stress
term in the momentum equation. The stiffness of the state equation is reduced
by using a value of $\gamma=1$ in the equation of state in contrast to the
traditionally chosen value of $\gamma=7$. The scheme produces excellent
results for internal flows and virtually eliminates particle clumping and void
regions. The scheme also displays reduced pressure oscillations. Recently, the
scheme has been extended to handle free-surface flows~\cite{zhang_hu_adams17}.

The Entropically Damped Artificially Compressible (EDAC) method of
Clausen~\cite{Clausen2013,Clausen2013a} is an alternative to the artificial
compressibility used by the weakly-compressible formulation. This method is
similar to the kinetically reduced local Navier-Stokes method presented in
\cite{krlns:ansumali:prl:2005,krlns:karlin:pre:2006,krlns:borok:pre:2007}.
However, the EDAC scheme uses the pressure instead of the grand potential as
the thermodynamic variable and this simplifies the resulting equations. The
EDAC scheme does not rely on an equation of state that relates pressure to
density. Instead, an evolution equation for the pressure is derived based on
thermodynamic considerations. The fluid is assumed to be isentropic and
minimization of density fluctuations leads to an equation for the pressure.
This equation includes a damping term for the pressure which reduces pressure
oscillations significantly. The scheme in its original form does not introduce
any new parameters into the simulation. There is also no need to introduce an
artificial viscosity in the momentum equation. It is important to note that
the EDAC method is based on artificial compressibility and therefore does
require the use of an artificial speed of sound. The method does therefore
have similar time step restrictions as the WCSPH scheme. The EDAC method was
validated for finite-difference~\cite{Clausen2013} and
finite-element~\cite{Clausen2013a} schemes and exhibited good parallel
performance owing to the elimination of the elliptic pressure Poisson
equation. Recently, Delorme et~al.~\cite{delorme_etal-17} have successfully
used the EDAC scheme in a high-order finite-difference solver with explicit
sub-grid-scale (SGS) terms for Large-Eddy-Simulation (LES).

In this work, the EDAC method is applied to SPH for the simulation of
incompressible fluids for both internal and free-surface problems. The
motivation for this work arose from the encouraging results (despite a
relatively naive implementation) presented in~\cite{PRKP:edac-sph-iccm2015}.
In that work, it was found that a simple application of the EDAC scheme
produced results that were better than the standard WCSPH, though not better
than those of the TVF scheme. Upon further investigation, it was found that
when the background pressure used in the TVF formulation is set to zero, the
EDAC scheme outperforms it. This is because the EDAC scheme provides a
smoother pressure distribution than that which is obtained via the equation of
state. There is no mechanism within the EDAC framework to ensure a uniform
distribution of particles. Therefore, we adapted the TVF scheme to be used
along with EDAC. The resulting scheme produces very good results and
outperforms the standard TVF for many of the benchmark problems considered in
this work.

The proposed EDAC scheme thus comes in two flavors. For internal flows, a
formulation based on the TVF is employed where a background pressure is added.
This background pressure ensures a homogeneous particle distribution. For
free-surface flows, a straight-forward formulation is used with the EDAC to
produce very good results. The scheme thus works well for both internal and
external flows. Several results are presented along with suitable comparisons
between the TVF and standard SPH schemes to demonstrate the new scheme. All
the results presented in this work are reproducible through the publicly
available PySPH package~\cite{PRKP:PySPH-particles13,PR:pysph:scipy16} along
with the code in \url{http://gitlab.com/prabhu/edac_sph}.

We note that the new scheme proposed is similar to the $\delta$-SPH
formulation of \citet{antuono-deltasph:cpc:2010} and
\citet{marrone-deltasph:cmame:2011}. The $\delta$-SPH scheme adds a
dissipation term to the continuity equation and uses a linearized equation of
state. The resulting scheme is very similar to the EDAC scheme. However, the
details of the implementation and origins of the scheme are different. In
addition, the present work uses the TVF formulation making the new scheme
considerably different in its final form.

The paper is organized as follows. In Section~\ref{sec:edac-scheme}, the
governing equations for the EDAC scheme are outlined. In
Section~\ref{sec:implementation}, the SPH discretization for the EDAC
equations are presented. In Section~\ref{sec:results}, the new scheme is
evaluated against a suite benchmark problems. The results are compared to the
analytical solution where available, and to the traditional WCSPH and TVF
formulations wherever possible. In Section~\ref{sec:conclusions}, the paper is
concluded with a summary.


\section{The EDAC method}
\label{sec:edac-scheme}

The EDAC method is discussed in detail in~\cite{Clausen2013,Clausen2013a}. In
this method, the density of the fluid $\rho$ is held fixed and an evolution
equation for the pressure based on thermodynamic considerations is derived. As
a result, a pressure evolution equation needs to be solved in addition to the
momentum equation. The equations are,

\begin{align}
    \label{eq:mom}
    \frac{d \ten{u}}{d t} &= -\frac{1}{\rho} \nabla p +
    \text{div}(\ten{\sigma}), \\
    \frac{d p}{d t} &= -\rho c_s^2 \text{div}(\ten{u}) + \nu
    \nabla^2 p,
    \label{eq:p-evolve}
\end{align}
where $\ten{u}$ is the velocity of the fluid, $p$ is the pressure, $\sigma$ is
the deviatoric part of the stress tensor, $c_s$ is the speed of sound, and
$\nu$ is the kinematic viscosity of the fluid.

As is typically chosen in WCSPH schemes, the speed of sound is set to a
multiple of the maximum fluid velocity. In this paper $c_s = 10\ u_{\max}$
unless otherwise mentioned.

In this work, the fluid is assumed to be Newtonian, which results in the
following momentum equation:

\begin{align}
    \label{eq:mom-newt}
    \frac{d \ten{u}}{d t} &= -\frac{1}{\rho} \nabla p +
    \nu \nabla^2 \ten{u}.
\end{align}

On comparison of the EDAC method with the standard WCSPH formulation, it can
be seen that the momentum equation is unchanged and
equation~\eqref{eq:p-evolve} replaces the continuity equation,
$\frac{d\rho}{dt} = -\rho \ \text{div}(\ten{u})$. Also, owing to the pressure
evolution equation in EDAC, there is no need for an equation of state to
couple the fluid density and pressure.

The EDAC equation~\eqref{eq:p-evolve}, is derived in \cite{Clausen2013}. Two
simplifying assumptions are made in this derivation. The first is to ignore
the viscous dissipation term, $\phi$. The second is to set the Prandtl number
$Pr = \gamma$. The first assumption is also made in the case of the
traditional artificial compressibility schemes (as can be seen in equation 7
of \citet{Clausen2013}). Clausen~\cite{Clausen2013} shows that the artificial
compressibility equation results when isentropic flow is assumed and this
implies that the fluid is inviscid and therefore the viscous dissipation is
ignored. For the EDAC equation, instead of assuming isentropy, one drives the
density fluctuations to zero resulting in a different thermodynamic
relationship. The viscous dissipation is neglected to simplify the resulting
equations. What is crucial in the EDAC equation is the second derivative of
pressure which smoothes the pressure through the introduction of entropy. The
second assumption to set $Pr =\gamma$ is arbitrary and in the numerical
simulations for the current work, a numerical viscosity value is found to be
more well suited (see section~\ref{sec:choice-of-nu}).

In summary, the EDAC scheme essentially introduces entropy by damping the
pressure oscillations. It is to be noted that this is the only difference from
the WCSPH. As mentioned in the introduction, a similar approach is used in
\cite{antuono-deltasph:cpc:2010,marrone-deltasph:cmame:2011}, where a damping
term is introduced in the continuity equation. With the EDAC scheme, the
pressure is evolved directly rather than computed from the density. It is
important to note that the introduction of pressure damping is not directly
equivalent to adding an artificial viscosity to the momentum equation as the
pressure only affects the momentum equation via its gradient. The approach is
similar to the density filtering approach (see
\cite{wcsph-state-of-the-art-2010} for more details). However, the pressure
field is smoothed at every step in the present case and not after every $m$
steps.

In the next section, an SPH-discretization of these equations is performed to
obtain the numerical scheme.

\section{Numerical implementation}
\label{sec:implementation}

As discussed in the introduction, there are two major issues that arise when
using weakly-compressible SPH (WCSPH) formulations. The first is the presence
of large pressure oscillations due to the stiff equation of state and the
second is due to the inhomogeneous particle distributions. The basic EDAC
formulation solves the first problem~\cite{PRKP:edac-sph-iccm2015}. The TVF
scheme solves the second problem by the introduction of a background pressure
for internal flows. Based on this, two different formulations using the EDAC
are presented in the following. The first formulation is what is called the
\emph{standard EDAC formulation}. This formulation can be used for external
flows. The second formulation is what is called the \emph{EDAC TVF
  formulation}, which is based on the TVF formulation and can be applied to
internal flows where it is possible to use a background pressure. Numerical
discretizations for both these schemes are discussed next.

\subsection{The standard EDAC formulation}
\label{sec:edac-ext}

The EDAC formulation keeps the density constant and this eliminates the need
for the continuity equation or the use of a summation density to find the
pressure. However, in SPH discretizations, $m/\rho$ is typically used as a
proxy for the particle volume. The density of the fluids can therefore be
computed using the summation density approach. This density does not directly
affect the pressure as there is no equation of state. In the case of solid
walls, the density of any wall particle is set to a constant. The classic
summation density equation for SPH is recalled:

\begin{equation}
  \label{eq:summation-density}
  \rho_i = \sum_j m_j W_{ij},
\end{equation}
where $W_{ij} = W(|\ten{r_i} - \ten{r_j}|, h)$ is the kernel function chosen
for the SPH discretization and $h$ is the kernel radius parameter. In this
paper, the quintic spline kernel is used, which is given by,

\begin{equation}
\label{eq:quintic-spline}
W(q) = \left \{
\begin{array}{ll}
  \alpha_2 \left[ {(3-q)}^5 - 6{(2-q)}^5 + 15{(1-q)}^5 \right],\
  & \textrm{for} \ 0\leq q \leq 1,\\
  \alpha_2 \left[ {(3-q)}^5 - 6{(2-q)}^5 \right],
  & \textrm{for} \ 1 < q \leq 2,\\
  \alpha_2 \ {(3-q)}^5 , & \textrm{for} \ 2 < q \leq 3,\\
  0, & \textrm{for} \ q>3,\\
\end{array} \right.
\end{equation}
where $\alpha_2 = 7/(478\pi h^2)$ in two-dimensions, and $q=|\ten{r}|/h$.

In the previous work~\cite{PRKP:edac-sph-iccm2015}, Monaghan's original
formulation was used for the pressure gradient and the formulation due to
\citet{morris-lowRe-97} was used for the viscous term in
equation~\eqref{eq:mom-newt}. The method of \citet{Adami2012} was used to
implement the effect of boundaries.

In the present work, a number density based formulation is employed as used in
\cite{Adami2012}, which results in the following momentum equation:

\begin{align}
  \label{eq:mom-sph}
  \frac{d \ten{u}_i}{d t} = \frac{1}{m_i} \sum_j \left(V_i^2 + V_j^2 \right)
  \left[ - \tilde{p}_{ij} \nabla W_{ij} + \tilde{\eta}_{ij}
    \frac{\ten{u}_{ij}}{(r_{ij}^2 + \eta h_{ij}^2)} \nabla W_{ij}\cdot
    \ten{r}_{ij} \right] + \ten{g}_i,
\end{align}
where $\ten{r}_{ij} = \ten{r}_i - \ten{r}_j$, $\ten{u}_{ij} = \ten{u}_i -
\ten{u}_j$, $h_{ij} = (h_i + h_j)/2$, $\eta=0.01$,

\begin{equation}
  \label{eq:number-density}
  V_i = \frac{1}{\sum_j W_{ij}},
\end{equation}
\begin{equation}
  \label{eq:tvf-p-ij}
  \tilde{p}_{ij} = \frac{\rho_j p_i + \rho_i p_j}{\rho_i + \rho_j},
\end{equation}
\begin{equation}
  \label{eq:tvf-eta-ij}
  \tilde{\eta}_{ij} = \frac{2 \eta_i \eta_j}{\eta_i + \eta_j},
\end{equation}
where $\eta_i = \rho_i \nu_i$.

The EDAC pressure evolution equation~(\ref{eq:p-evolve}) is discretized using
a similar approach to the momentum equation to be,

\begin{align}
    \label{eq:p-edac}
    \frac{d p_i }{dt} &=  \sum_j \frac{m_j \rho_i}{\rho_j} c_s^2 \ \ten{u_{ij}}
    \cdot \nabla W_{ij} + \frac{(V_i^2 + V_j^2)}{m_i} \tilde{\eta}_{ij}
    \frac{p_{ij}}{(r_{ij}^2 + \eta h_{ij}^2)} \nabla W_{ij}\cdot \ten{r}_{ij},
\end{align}
where $p_{ij} = p_i - p_j$.  The particles are moved according to,

\begin{align}
  \label{eq:motion}
  \frac{d \ten{r}_i}{d t} = \ten{u}_i.
\end{align}

Upon the specification of suitable initial conditions for $\ten{u}$, $p$, $m$,
and $\ten{r}$,
equations~\eqref{eq:summation-density},~\eqref{eq:mom-sph},~\eqref{eq:p-edac},
and~\eqref{eq:motion} are sufficient for simulating the flow in the absence of
any boundaries.

\subsection{EDAC TVF formulation}
\label{sec:edac-int}

In WCSPH, as the particles move they tend to become disordered. This
introduces significant errors in the simulation. The particle positions can be
regularized by the addition of a background pressure. A naive approach would
be to simply add a constant pressure and use it in the governing equations.
However, as shown by \citet{sph:basa-etal-2009}, the error in computing the
gradient of pressure increases when the pressure values are large. They
subtract the average pressure to reduce this error. The TVF scheme of
\citet{Adami2013} overcomes this by advecting the particles using an arbitrary
background pressure through the ``transport velocity'' and correct for this
background pressure using an additional stress term in the momentum equation.
This ensures a homogeneous particle distribution without introducing a
constant background pressure in the pressure derivative term.

For internal flows, the TVF formulation is adapted to introduce the background
pressure. The density is computed using the summation density
equation~(\ref{eq:summation-density}). As before, this is mainly to serve as a
proxy for the particle volume in the SPH discretizations. The momentum
equation for the TVF scheme as discussed in \citet{Adami2013} is given by,

\begin{equation}
  \label{eq:tvf-momentum}
  \begin{split}
    \frac{\tilde{d} \ten{u}_i}{d t} = \frac{1}{m_i} \sum_j \left( V_i^2 +
    V_j^2 \right) & \left[ - \tilde{p}_{ij} \nabla W_{ij} +
      \frac{1}{2}(\ten{A}_i + \ten{A}_j) \cdot \nabla W_{ij} \right . \\ &
      \left .  + \tilde{\eta}_{ij} \frac{\ten{u}_{ij}}{(r_{ij}^2 + \eta
        h_{ij}^2)} \nabla W_{ij}\cdot \ten{r}_{ij} \right] + \ten{g}_i,
  \end{split}
\end{equation}
where $\ten{A} = \rho \ten{u}(\ten{\tilde{u}} - \ten{u})$, $\ten{\tilde{u}}$
is the advection or transport velocity and the material derivative,
$\frac{\tilde{d}}{dt}$ is given as,

\begin{align}
    \label{eq:tvf-derivative}
    \frac{\tilde{d}( \cdot) }{d t} = \frac{\partial (\cdot) }{\partial t} +
  \ten{\tilde{u}} \cdot \text{grad} (\ten{\cdot}).
\end{align}
Thus the particles move using the transport velocity,

\begin{equation}
  \label{eq:advection}
  \frac{d \ten{r}_i}{dt} = \ten{\tilde{u}}_i.
\end{equation}
The transport velocity is obtained from the momentum velocity $\ten{u}$ at
each time step using,

\begin{equation}
  \label{eq:transport-vel}
  \ten{\tilde{u}}_i(t + \delta t) = \ten{u}_i(t) +
  \delta t \left(
    \frac{\tilde{d} \ten{u}_i}{dt}
    - \frac{p_b}{m_i} \sum_j \left( V_i^2 + V_j^2 \right)
    \nabla W_{ij}
    \right),
\end{equation}
where $p_b$ is the background pressure.

In the TVF scheme, the pressure is computed from the density using the
standard equation of state with a value of $\gamma=1$. Instead, the EDAC
equation~(\ref{eq:p-edac}) is used to evolve the pressure. In the present
approach, the pressure reduction technique proposed by
\citet{sph:basa-etal-2009} is used to mitigate the errors due to large
pressures. This requires the computation of the average pressure of each
particle, $p_{\text{avg}}$:

\begin{equation}
  \label{eq:pavg}
  p_{\text{avg}, i} = \sum_{j=1}^{N_i} \frac{p_j}{N_i},
\end{equation}
where $N_i$ are the number of neighbors for the particle $i$ and includes both
fluid and boundary neighbors. Equation~(\ref{eq:tvf-p-ij}) is then replaced
with,

\begin{equation}
  \label{eq:tvf-p-ij-basa}
  \tilde{p}_{ij} =
  \frac{\rho_j (p_i-p_{avg, i}) + \rho_i (p_j - p_{avg, i})}{\rho_i + \rho_j}.
\end{equation}

In Section~\ref{sec:results} it can be seen that this results in significantly
improved results that outperform the traditional TVF scheme. It is worth
mentioning that this technique, applied to the standard SPH or to the standard
TVF scheme does not result in any significant improvement.

The boundary conditions are satisfied using the formulation of
\citet{Adami2012}. This method uses fixed wall particles and sets the pressure
and velocity of these wall particles in order to accurately simulate the
boundary conditions. The same scheme is used here with the only modification
being that the density of the boundary particles is not set based on the
pressure of the boundary particles (i.e.\ equation~(28) in \citet{Adami2012}
is not used).

\subsection{Suitable choice of $\nu$ for EDAC}
\label{sec:choice-of-nu}

In equation~\eqref{eq:p-edac} one can see that the viscosity $\nu$ is used to
diffuse the pressure. The original formulation assumes that the value of $\nu$
is the same as the fluid viscosity. In our numerical experiments it was found
that if the viscosity is too small, the pressure builds up too fast and
eventually blows up. If the viscosity is too large it diffuses too fast
resulting in a non-physical simulation. Thus, the physical viscosity is not
always the most appropriate. Instead using,

\begin{equation}
  \label{eq:edac-nu}
  \nu_{edac} = \frac{\alpha h c_s}{8},
\end{equation}
works very well. The choice of $\nu_{edac}$ is motivated by the expression for
artificial viscosity in traditional WCSPH formulations. The form of the
viscous term used in
\cite{antuono-deltasph:cpc:2010,marrone-deltasph:cmame:2011} is also the same.
In this paper, it is found that $\alpha=0.5$ is a good choice for a wide range
of Reynolds numbers (0.0125 to 10000). While this choice of $\nu_{edac}$ is
motivated by the expression for artificial viscosity traditionally used in the
SPH, the viscous damping of pressure is not the same as adding artificial
viscosity directly to the momentum equation.

To summarize the schemes,
\begin{itemize}
\item for external flow problems,
  equations~\eqref{eq:summation-density},~\eqref{eq:mom-sph},
  and~\eqref{eq:p-edac} are used. The particles move with the fluid velocity
  $\ten{u}$ and are advected according to~\eqref{eq:motion}.

\item for internal flows,
  equations~\eqref{eq:summation-density},~\eqref{eq:tvf-momentum},~\eqref{eq:pavg},
  \eqref{eq:tvf-p-ij-basa} and~\eqref{eq:p-edac} are used.
  Equation~\eqref{eq:advection} is used to advect the particles. The transport
  velocity is found from equation~\eqref{eq:transport-vel}.
\end{itemize}
For each of the schemes, the value of $\nu$ used in the
equation~\eqref{eq:p-edac} is found using equation~\eqref{eq:edac-nu}. The
value of $\nu$ used in the momentum equation is the fluid viscosity.

The proposed EDAC scheme is explicit and as such, any suitable integrator can
be used. In this work, one of the two simplest possible two-stage explicit
integrators is chosen. For both integrators, the particle properties are first
predicted at $t + \delta t/2$. The right-hand-side (RHS) is subsequently
evaluated at this intermediate step and the final properties at $t + \delta t$
are obtained by correcting the predicted values. Two variants of this
predictor-corrector integration scheme are defined. In the first type, the
prediction stage is completed using the RHS from the previous time-step. This
is called the Predict-Evaluate-Correct (PEC) type integrator. In the second
variant, an evaluation of the RHS is carried out for the predictor stage. This
integrator, deemed Evaluate-Predict-Evaluate-Correct (EPEC) is more accurate
(at the cost of two RHS evaluations per time-step).  For the standard EDAC
scheme the time integration proceeds as follows.  The predictor step is first
performed as,

\begin{equation}
  \label{eq:integrate-edac-pred}
  \begin{aligned}
  \ten{u}^{n+\frac{1}{2}} &= \ten{u}^n + \frac{\Delta t}{2m} \ten{f}^{n-\frac{1}{2}} \\
  \ten{r}^{n+\frac{1}{2}} &= \ten{r}^n + \frac{\Delta t}{2} \ten{u}^{n-\frac{1}{2}} \\
  p^{n+\frac{1}{2}} & = p^n + \frac{\Delta t}{2}\ a_p^{n-\frac{1}{2}},
\end{aligned}
\end{equation}

\noindent where $a_p$ is the right hand side of equation~\eqref{eq:p-edac}. The new
accelerations are then computed at this point and the corrector step is as,

\begin{equation}
  \label{eq:integrate-edac-corr}
  \begin{aligned}
  \ten{u}^{n+1} &= \ten{u}^n + \frac{\Delta t}{m} \ \ten{f}^{n+\frac{1}{2}} \\
  \ten{r}^{n+1} &= \ten{r}^n + \frac{\Delta t}{m} \ \ten{u}^{n+\frac{1}{2}} \\
  p^{n+1} & = p^n + \Delta t \ a_p^{n+\frac{1}{2}}.
\end{aligned}
\end{equation}

\noindent For the EDAC TVF method, the predictor step is implemented as,

\begin{equation}
  \label{eq:integrate-edac-tvf-pred}
  \begin{aligned}
    \ten{u}^{n+\frac{1}{2}} &= \ten{u}^n + \frac{\Delta t}{2m} \  \ten{f}^{n-\frac{1}{2}} \\
    \ten{\tilde{u}}^{n+\frac{1}{2}} &= \ten{u}^{n+\frac{1}{2}} +
    \frac{\Delta t}{2m}  \ \ten{f_{p_b}}^{n-\frac{1}{2}} \\
  \ten{r}^{n+\frac{1}{2}} &= \ten{r}^n + \frac{\Delta t}{2} \  \ten{\tilde{u}}^{n+\frac{1}{2}} \\
  p^{n+\frac{1}{2}} & = p^n + \frac{\Delta t}{2}\ a_p^{n-\frac{1}{2}},
\end{aligned}
\end{equation}

\noindent where $\ten{f}_{p_b}$ is the background pressure force.  At this point the
accelerations are computed and the corrector step is performed as,

\begin{equation}
  \label{eq:integrate-edac-tvf-corr}
  \begin{aligned}
  \ten{u}^{n+1} &= \ten{u}^n + \frac{\Delta t}{m} \  \ten{f}^{n+\frac{1}{2}} \\
  \ten{\tilde{u}}^{n+1} &= \ten{u}^{n+1} + \frac{\Delta t}{m} \  \ten{f_{p_b}}^{n+\frac{1}{2}} \\
  \ten{r}^{n+1} &= \ten{r}^n + \Delta t \ \ten{\tilde{u}}^{n+1} \\
  p^{n+1} & = p^n + \Delta t \ a_p^{n+\frac{1}{2}}.
\end{aligned}
\end{equation}

As mentioned in the introduction, all the equations and algorithms presented
in this work are implemented using the PySPH
framework~\cite{PRKP:PySPH-particles13,PR:pysph:scipy16,pysph}. PySPH is an
open source framework for SPH that is written in Python. It is easy to use,
easy to extend, and supports non-intrusive parallelization and dynamic load
balancing. PySPH provides an implementation of the TVF formulation and this
allows for a comparison of the results with those of the standard SPH and TVF
where necessary. In the next section, the performance of the proposed SPH
scheme is evaluated for several benchmark problems of varying complexity.


\section{Numerical Results}
\label{sec:results}

In this section the EDAC scheme is applied to a suite of test problems. The
results from the new EDAC scheme are compared with the standard weakly
compressible SPH (WCSPH) and, where possible, with those from the
Transport-Velocity-Formulation (TVF) scheme~\cite{Adami2013}.

Every attempt has been made to allow easy reproduction of all of the present
results. The TVF implementation is available as part of PySPH~\cite{pysph} as
is an implementation of EDAC-SPH. Every figure in this article is
automatically generated. The approach and tools used for this are described in
detail in \cite{pr:automan:2018}. The code for the EDAC implementation and the
automation of all of our results are available from
\url{http://gitlab.com/prabhu/edac_sph}.

\subsection{Taylor Green Vortex}
\label{sec:tgv}

The Taylor-Green vortex problem is a particularly challenging case to simulate
using SPH. This is an exact solution of the Navier-Stokes equations in a
periodic domain. Here, a two-dimensional version is considered as is done in
\cite{Adami2013}. The fluid is considered periodic in both directions and the
exact solution is given by,

\begin{align}
  \label{eq:tgv_sol}
  u &= - U e^{bt} \cos(2 \pi x) \sin(2 \pi y) \\
  v &=   U e^{bt}\sin(2 \pi x) \cos(2 \pi y) \\
  p &=  -U^2 e^{2bt} (\cos(4 \pi x) + \cos(4 \pi y))/4,
\end{align}
where $U$ is chosen as $1m/s$, $b=-8\pi^2/Re$, $Re=U L /\nu$, and $L=1m$.

The Reynolds number, $Re$, is initially chosen to be $100$. The flow is
initialized with $u, v, p$ set to the values at $t=0$. The evolution of the
quantities are studied for different numerical schemes. The speed of sound is
set to $10$ times the maximum flow velocity at $t=0$. The background pressure
is set as discussed by \citet{Adami2013} to $p_b = c_s^2 \rho$. The quintic
spline kernel is used with the smoothing length $h$ set to the particle
spacing $\Delta x$. The value of $\alpha$ in the equation~\eqref{eq:edac-nu}
is chosen as 0.5. The results from the standard SPH scheme, the TVF, and the
new scheme are compared. Since a physical viscosity is used and the solution
to the problem remains smooth, no artificial viscosity is used for any of the
schemes.

A Predict-Evaluate-Correct (PEC) integrator with a fixed time-step is used and
chosen as per the following equation,

\begin{equation}
  \label{eq:adaptive_timestep}
  \Delta t = \min\left( \frac{h}{4(c_s+|U_{max}|)},\  \frac{h^2}{8\nu}\right).
\end{equation}
Unless explicitly mentioned, all simulations use this integrator and a
time-step chosen as above.

In \citet{Adami2013}, the simulation starts with either uniformly distributed
particles or with a ``relaxed initial condition''. For the relaxed initial
condition, the authors use the particle distribution generated by the
uniformly distributed case at the final time and impose an analytical initial
condition at the particle positions. The results for the uniformly distributed
particles have about an order of magnitude more error than that of the relaxed
initialization. This is because the uniform distribution results in particles
being placed along (or near) stagnation streamlines resulting in non-uniform
particle distributions.

In this work, for this particular problem, the initial distribution is uniform
but a small random displacement is added to the particles. The random
displacement is uniformly distributed and the maximum displacement in any
coordinate direction is chosen to be $\Delta x/5$. The same initial conditions
are used for all schemes. This is simple to implement, resolves the problems
with stagnation streamlines, and enables for a fair comparison of all the
schemes.

\begin{figure}[htpb]
\centering
\includegraphics[width=10cm]{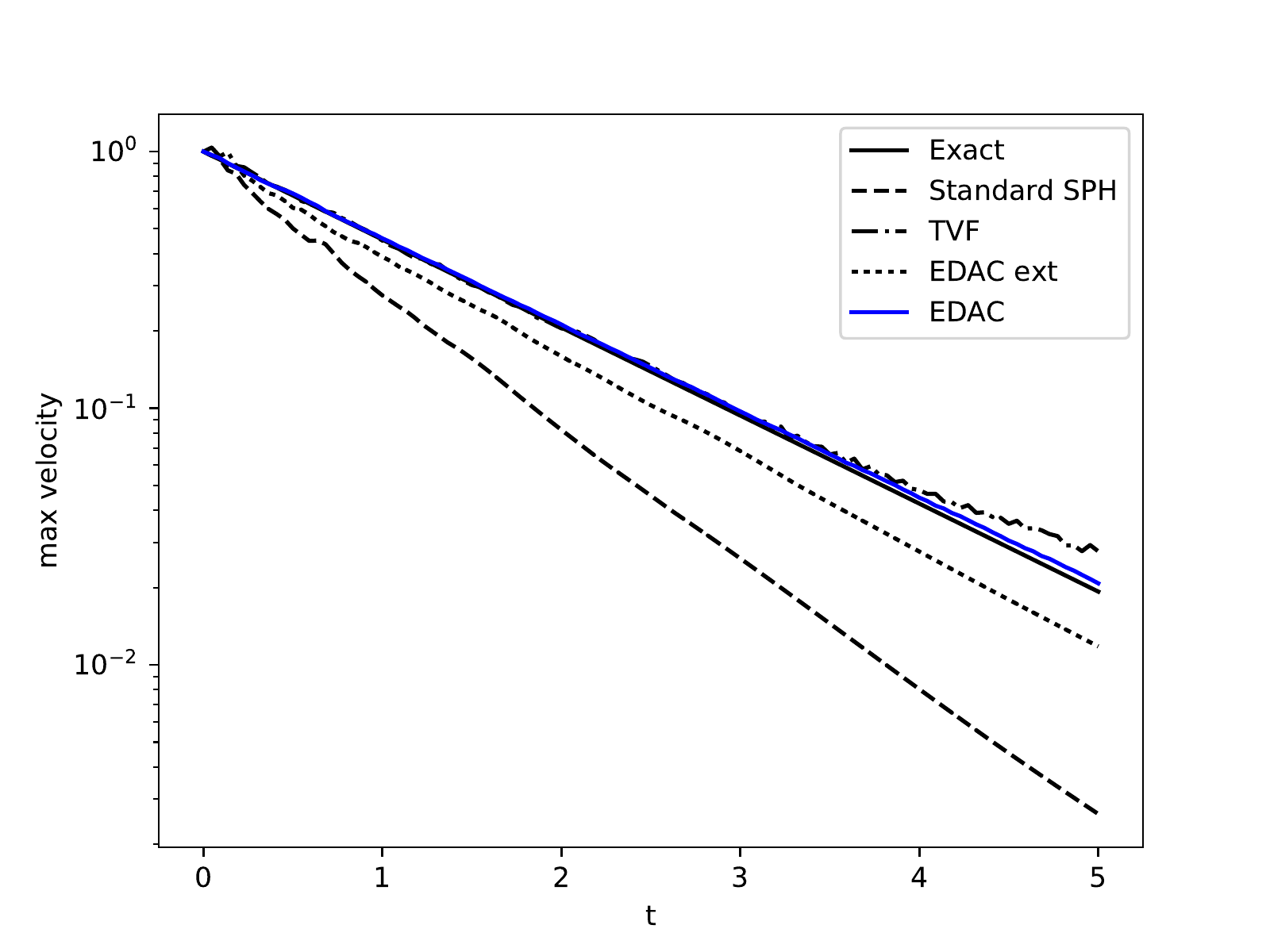}
\caption{The decay with time of the velocity magnitude for the different
  schemes. Particles are initialized with $n_x = n_y = 50$ and thereafter
  randomly perturbed. The Reynold's number is chosen to be $Re=100$. The
  quintic spline kernel is used with a smoothing length equal to the initial
  (undisturbed) particle spacing.}
\label{fig:tgv:decay}
\end{figure}
In Fig.~\ref{fig:tgv:decay}, the decay of the maximum velocity magnitude
produced by different schemes is compared with the exact solution. A regular
particle distribution with $n_x = n_y = 50$ is randomly perturbed as discussed
above. The standard SPH, TVF, standard EDAC (labeled EDAC ext), and TVF EDAC
(labeled EDAC) schemes are compared. As can be seen, the EDAC and TVF perform
best. The standard EDAC without the TVF (labeled EDAC ext) is better than the
standard SPH but not as effective as the TVF scheme. As discussed in previous
sections, this occurs because the TVF background pressure results in a more
homogeneous particle distribution.

\begin{figure}[htpb]
\centering
\includegraphics[width=10cm]{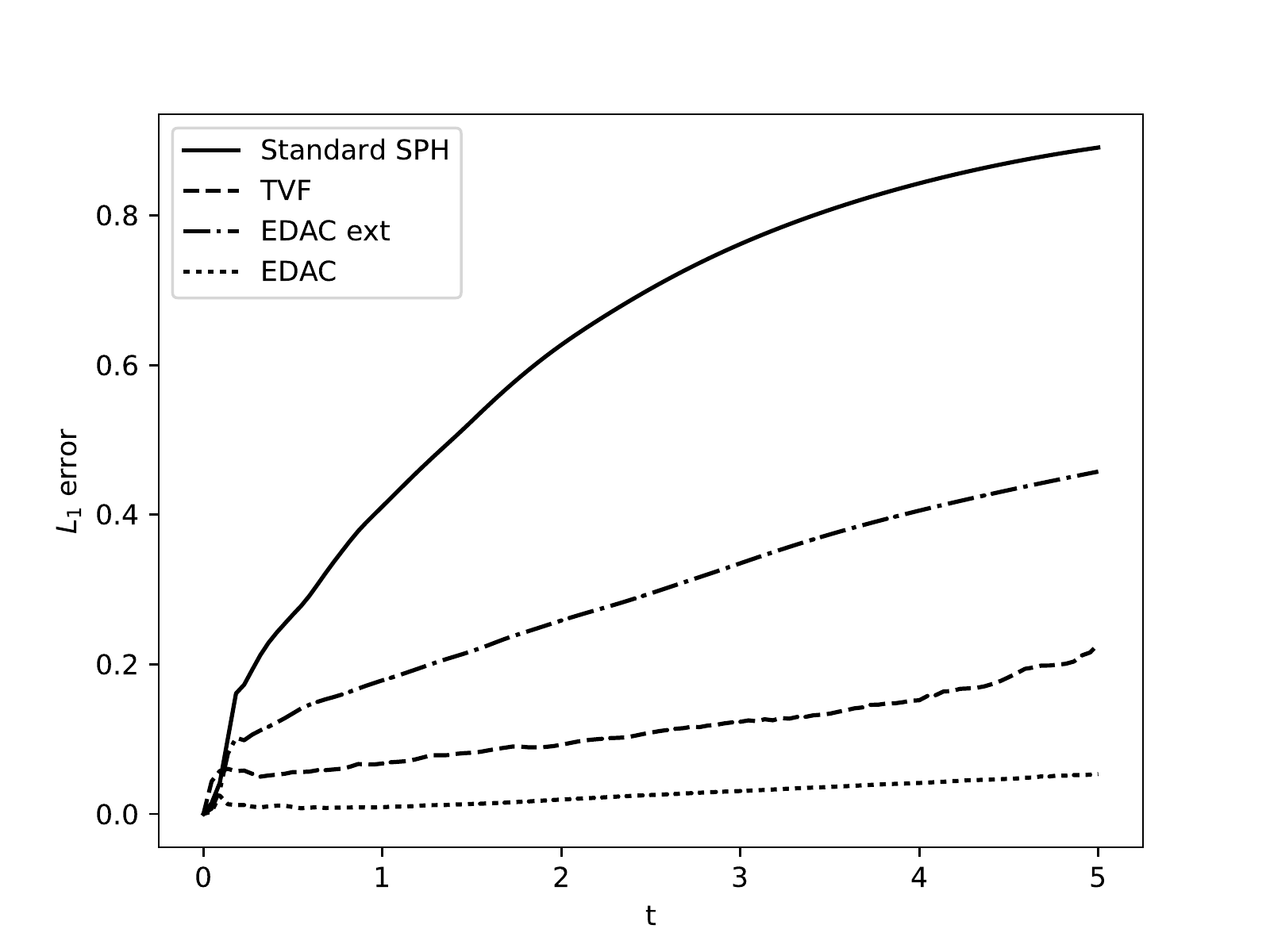}
\caption{The $L_1$ error of the velocity magnitude vs.\ $t$ for the
  standard SPH (solid line), standard EDAC (dash-dot), EDAC TVF (dot) and TVF
  (dash) schemes.}
\label{fig:tgv:l1_error}
\end{figure}

Fig.~\ref{fig:tgv:decay} does not clearly differentiate between schemes. The
$L_1$ error of $|\ten{u}|$ is a better measure of the performance of the
schemes and is plotted in Fig.~\ref{fig:tgv:l1_error}. The $L_1$ error is
computed as the average value of the difference between the exact velocity
magnitude and the computed velocity magnitude, that is,

\begin{equation}
  \label{eq:tgv_l1}
  L_1 = \frac{\sum_i |\ten{u_{i, computed}}| - |\ten{u_{i, exact}}|}
  {\sum_i |\ten{u_{i, exact}}|},
\end{equation}
where the value of $\ten{u}$ is computed at the particle positions for each
particle $i$ in the flow.

Fig.~\ref{fig:tgv:l1_error} clearly brings out the differences in the schemes.
It is easy to see that the TVF EDAC scheme (labeled EDAC) produces much lower
errors than the TVF scheme (by almost a factor of 4). The difference between
the standard EDAC scheme and the TVF is also brought out. It is easy to see
that the standard EDAC scheme (labeled EDAC ext) is better than the standard
SPH.

\begin{figure}
\centering
\includegraphics[width=10cm]{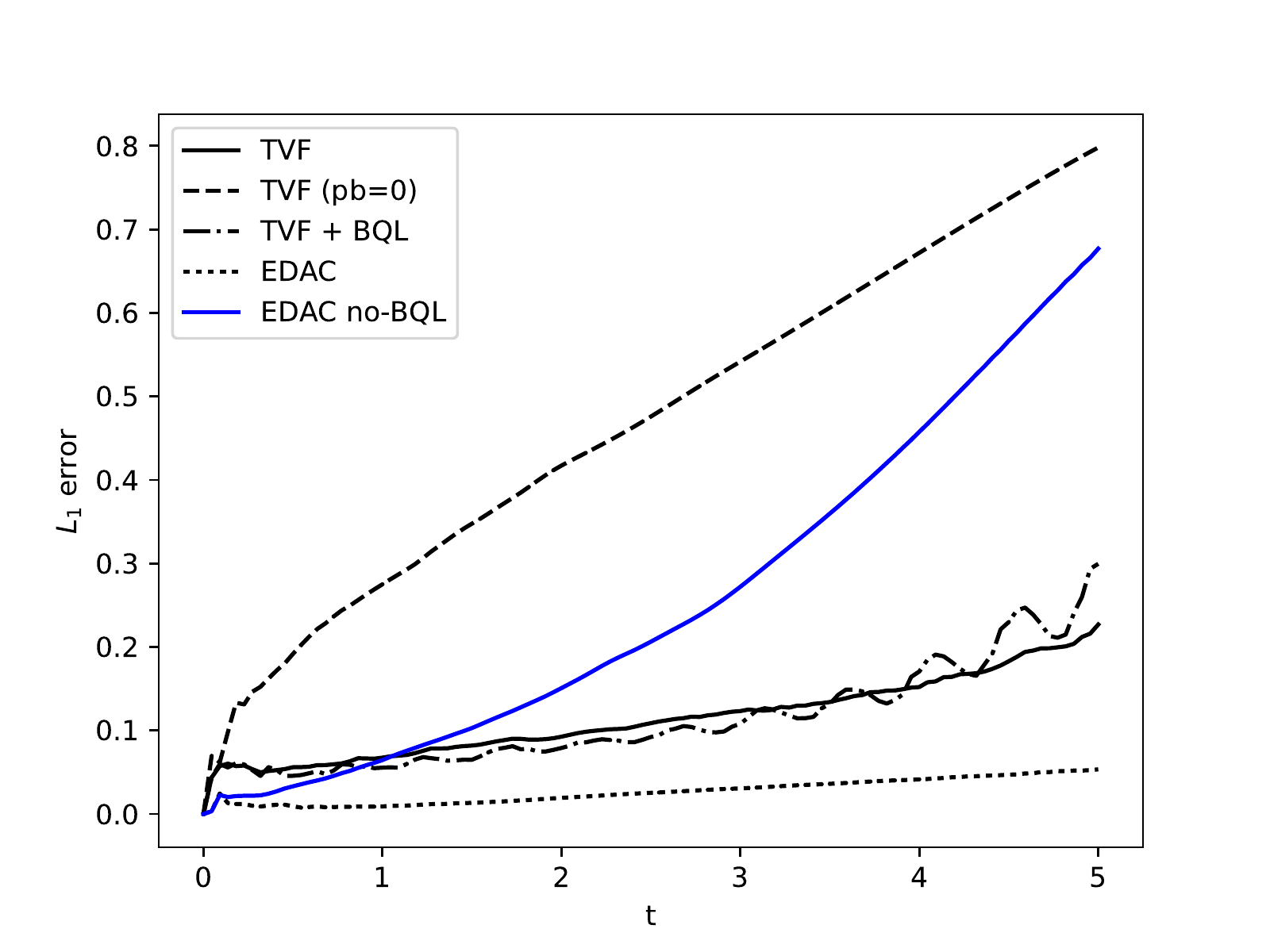}
\caption{The $L_1$ error of the velocity magnitude versus $t$ for
  other variations of the schemes.}
\label{fig:tgv:l1_error_other}
\end{figure}

In order to better understand the behavior of the methods, several other
variations of the basic schemes have been studied.
Fig.~\ref{fig:tgv:l1_error_other} shows the $L_1$ error of the velocity
magnitude using the TVF formulation, along with the background pressure
correction scheme of \citet{sph:basa-etal-2009} (labeled as ``TVF + BQL'').
The results of using the TVF without any background pressure is labeled as
``TVF (pb=0)''. This clearly shows that the pressure correction of Basa et
al.\ does not affect the TVF scheme, and that without the background pressure,
the standard EDAC is in fact better than the TVF. While this is only to be
expected, it does highlight that the EDAC scheme performs very well. The plot
labeled ``EDAC no-BQL'' demonstrates that the pressure correction due to
\citet{sph:basa-etal-2009} is necessary for the EDAC scheme. It is also
found (not shown here) that using the pressure correction with the standard
SPH formulation does not produce any significant advantages. Similarly, the
tensile correction of Monaghan~\cite{sph:tensile-instab:monaghan:jcp2000} has
no major influence on the results.

\begin{figure}
\centering
\includegraphics[width=10cm]{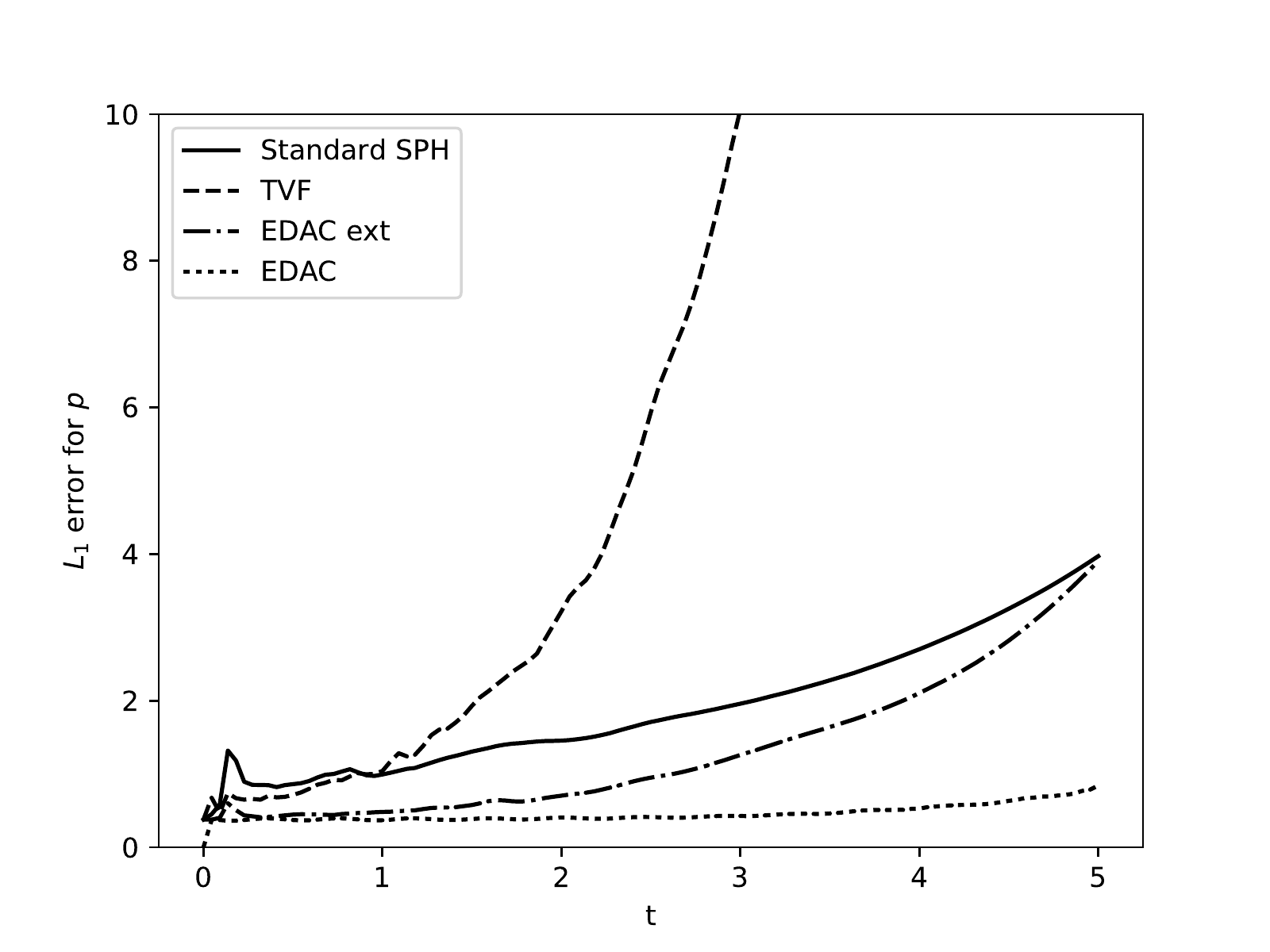}
\caption{The $L_1$ error of the pressure versus $t$ for the Standard
  SPH (solid line), TVF (dash), EDAC-ext (dash-dot) and EDAC (dot)
  schemes.}
\label{fig:tgv:p_l1_error}
\end{figure}

The EDAC scheme evolves the pressure in a very different manner from the
traditional WCSPH schemes. It is important to see how it captures the pressure
field as compared with the other schemes. In Fig.~\ref{fig:tgv:p_l1_error},
the $L_1$ error in the pressure is plotted as the simulation evolves. The
pressure in the EDAC scheme drifts due to the use of the transport velocity
used to move the particles, we therefore compute $p-p_{avg}$ where $p_{avg}$
is computed using equation~\eqref{eq:pavg}. In order to make the comparisons
uniform this is done for all the schemes. This does not change the quality of
the results by much. The error is computed as,

\begin{equation}
  \label{eq:tgv_p_l1}
  p_{L_1} = \frac{\sum_i |p_{i, computed} - p_{i, avg} - p_{i, exact}|}
  {\text{max}_{i} (p_{i, exact})}.
\end{equation}

As can be clearly seen in Fig.~\ref{fig:tgv:p_l1_error}, the new EDAC scheme
outperforms all other schemes.  It should be noted that the magnitude of the
error in the pressure is rather large for all schemes suggesting that all of
the schemes have difficulty in capturing the pressure field accurately.

\begin{figure}
\centering
\includegraphics[width=10cm]{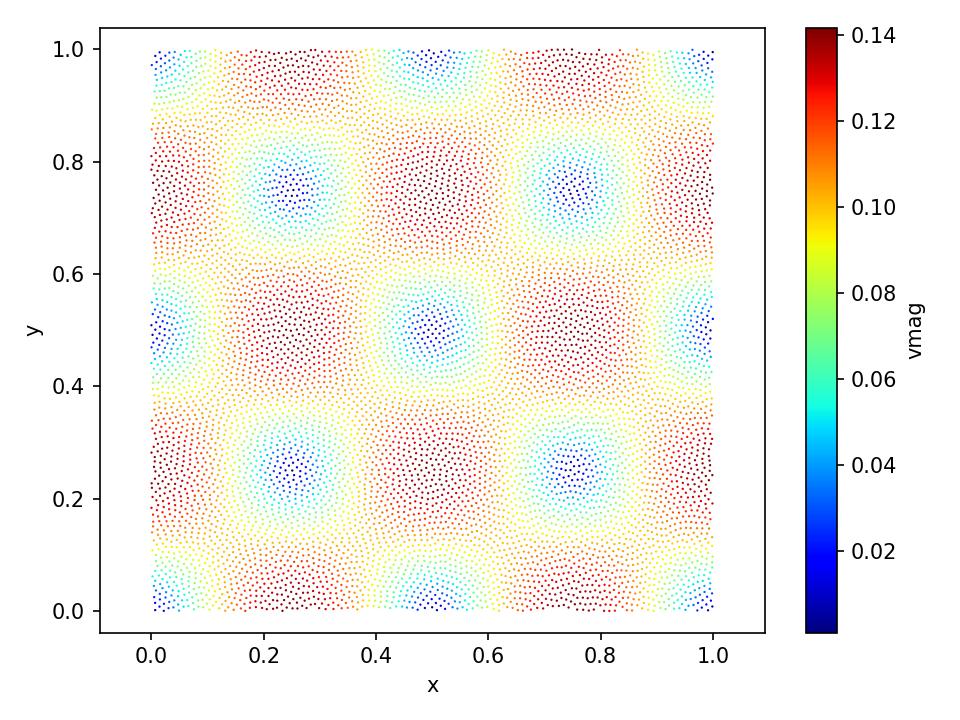}
\caption{The distribution of particles at $t=2.5$ for the simulation using the
  EDAC scheme with $n_x=100$. The colors indicate the velocity magnitude.}
\label{fig:tgv:edac-vmag}
\end{figure}

\begin{figure}
  \centering
  \begin{subfigure}[b]{0.48\textwidth}
    \includegraphics[width=\textwidth]{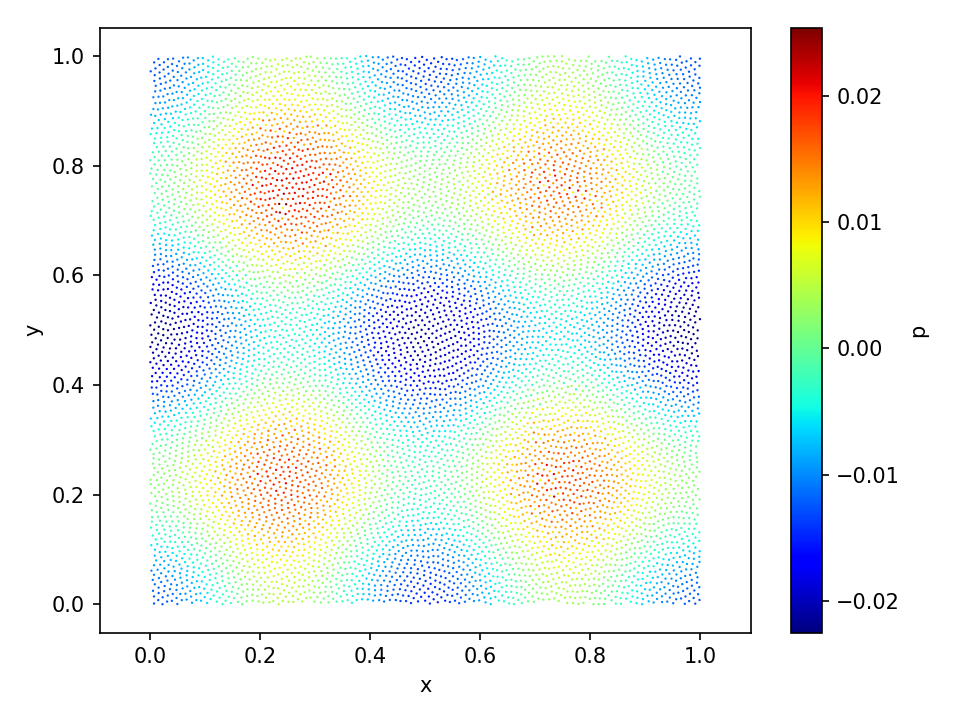}
    \caption{EDAC scheme.}
    \label{fig:tgv:edac-p}
  \end{subfigure}
  ~
  \begin{subfigure}[b]{0.48\textwidth}
    \includegraphics[width=\textwidth]{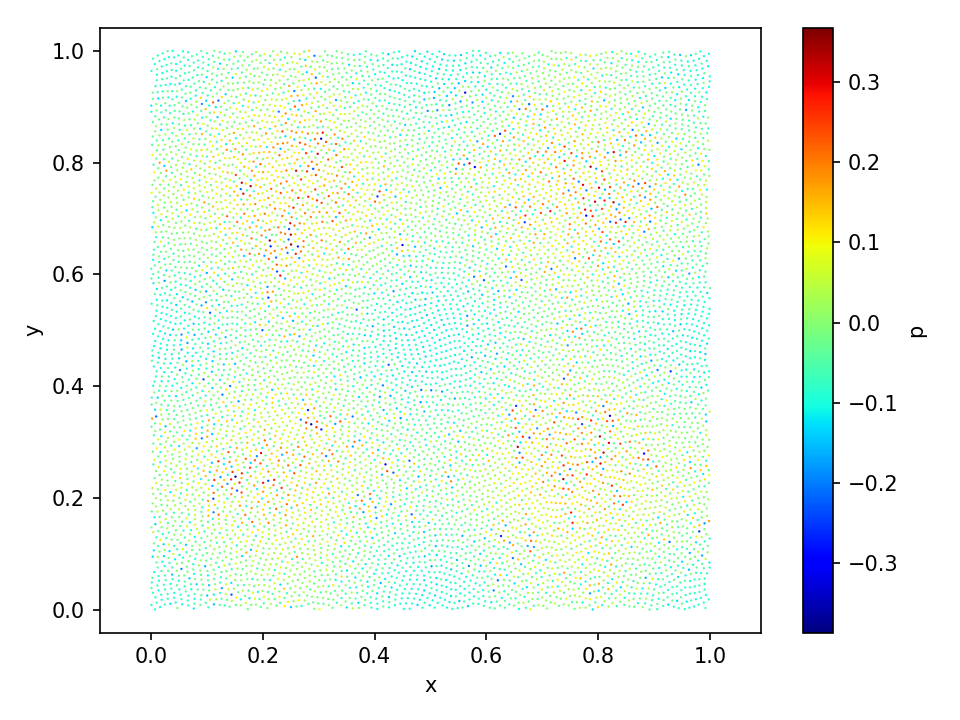}
    \caption{TVF scheme.}
    \label{fig:tgv:tvf-p}
  \end{subfigure}

  \caption{The distribution of particles showing the pressure at $t=2.5$ with
    $n_x=100$. The left column shows the results using the EDAC and the right
    show the results with the TVF scheme.}
\label{fig:tgv:p-compare}
\end{figure}

Fig.~\ref{fig:tgv:edac-vmag} shows the distribution of particles for the case
where $n_x=100$ using the EDAC scheme at a time of $t=2.5$. The color
indicates the velocity magnitude. As can be seen, the particles are
distributed homogeneously and the results are good. Fig.~\ref{fig:tgv:edac-p}
shows the distribution of particles with the color indicating the pressure.
The pressure plotted is the difference of the local pressure minus the average
of the pressure of all particles. Fig.~\ref{fig:tgv:tvf-p} shows the same
results when obtained using the TVF. The results show clearly that the EDAC
scheme performs well and this is consistent with the $L_1$ errors of pressure
shown in Fig.~\ref{fig:tgv:p_l1_error}.

\begin{figure}
\centering
\includegraphics[width=10cm]{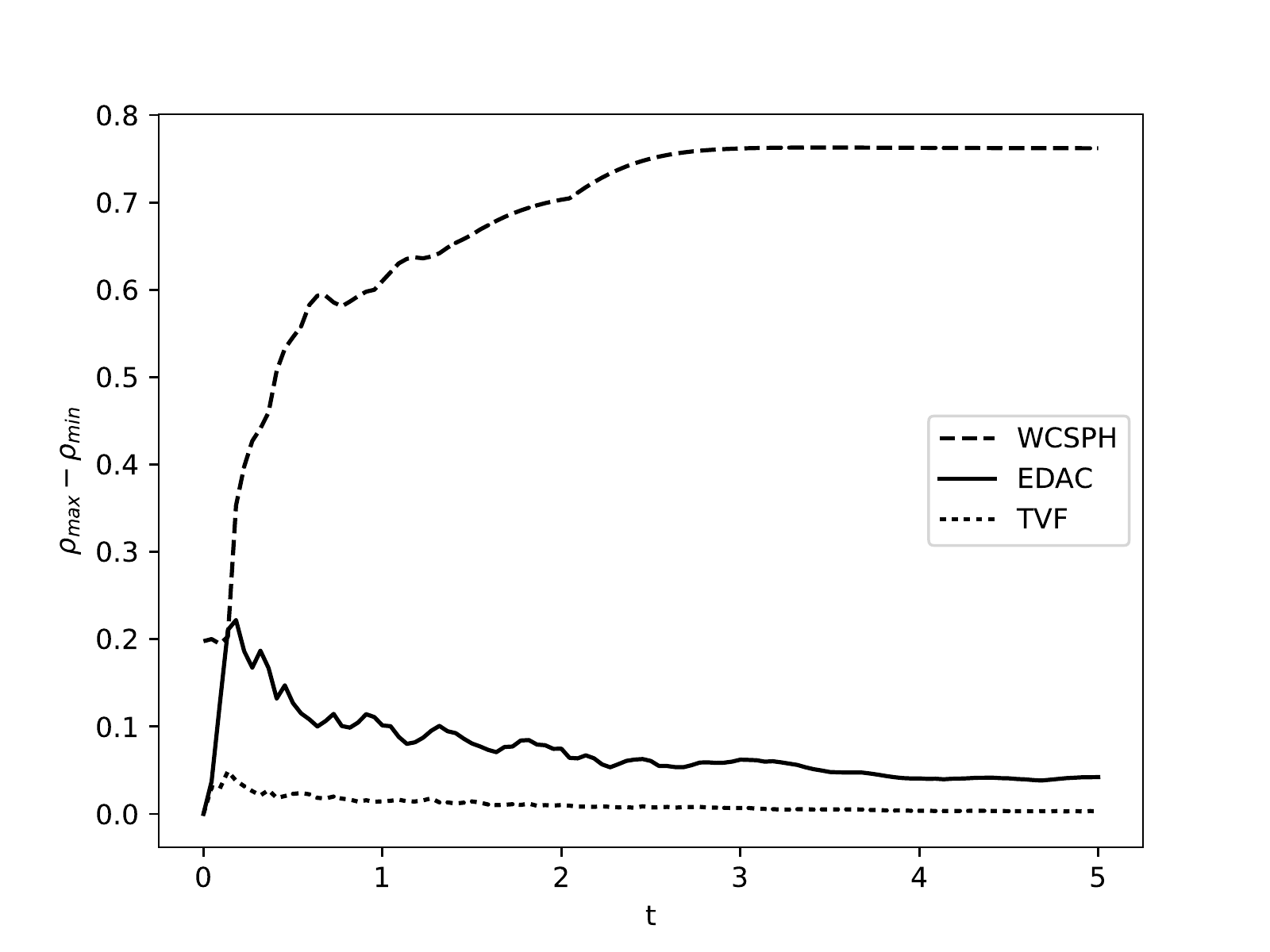}
\caption{The variation in the density computed using the summation density
  versus time for the WCSPH scheme, the EDAC scheme, and the TVF. The number
  of particles are $n_x=50$ and $Re=100$. }
\label{fig:tgv:density-variation}
\end{figure}

In Fig.~\ref{fig:tgv:density-variation}, the variation in density computed
using the summation density is plotted versus time. The Reynolds number is
chosen as 100 and $n_x=50$ for all schemes. The variation is computed as the
difference between the maximum and minimum density at that time. The density
is computed using the summation density for all the schemes. The WCSPH scheme
uses a continuity equation to evolve the density. The EDAC scheme only evolves
the pressure. The plot serves as a test of how well the volume is preserved by
the schemes. The EDAC scheme is far superior to the WCSPH scheme. The TVF
scheme uses the summation density to compute the density and displays the
smallest density variations.

\begin{figure}
\centering
\includegraphics[width=10cm]{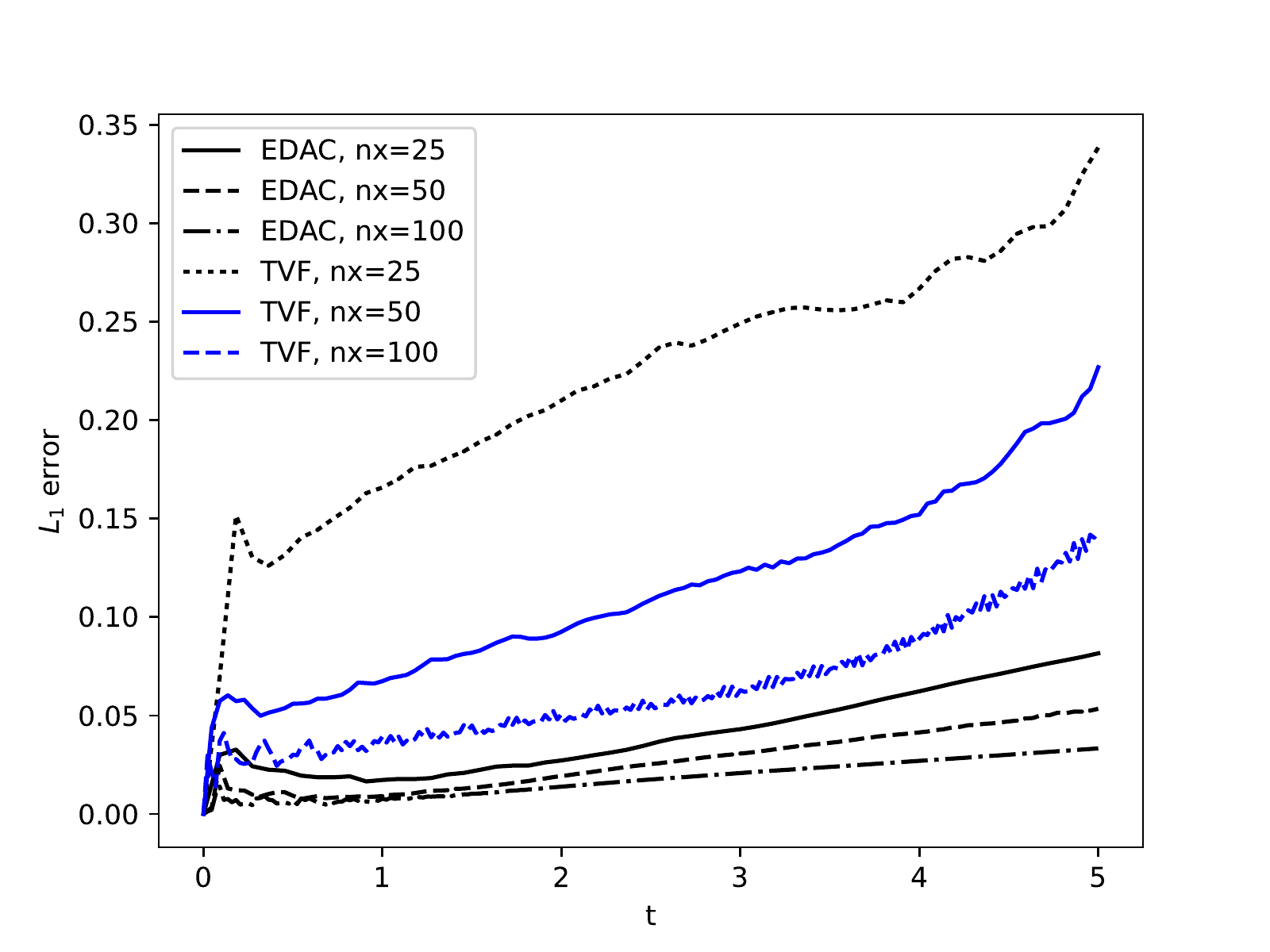}
\caption{The $L_1$ error of the velocity magnitude versus $t$ for different
  resolutions.}
\label{fig:tgv:conv}
\end{figure}

In Fig.~\ref{fig:tgv:conv}, the $L_1$ error for the velocity magnitude is
plotted but for different values of the initial particle spacing $n_x$. We
note that $n_x =25$ corresponds to a $\Delta x = 0.04$. As can be seen, the
EDAC scheme (Section~\ref{sec:edac-int}) consistently produces less error than
the TVF scheme at even such low resolutions.

From Fig.~\ref{fig:tgv:conv} it can be seen that with just $25\times25$
particles, the EDAC produces about 3 times less error than the TVF. It is to
be noted that for this low resolution, the random initial perturbation of the
particles is limited to a maximum of $\Delta x/10$ instead of the $\Delta x/5$
for the other cases.

\begin{figure}
\centering
\includegraphics[width=10cm]{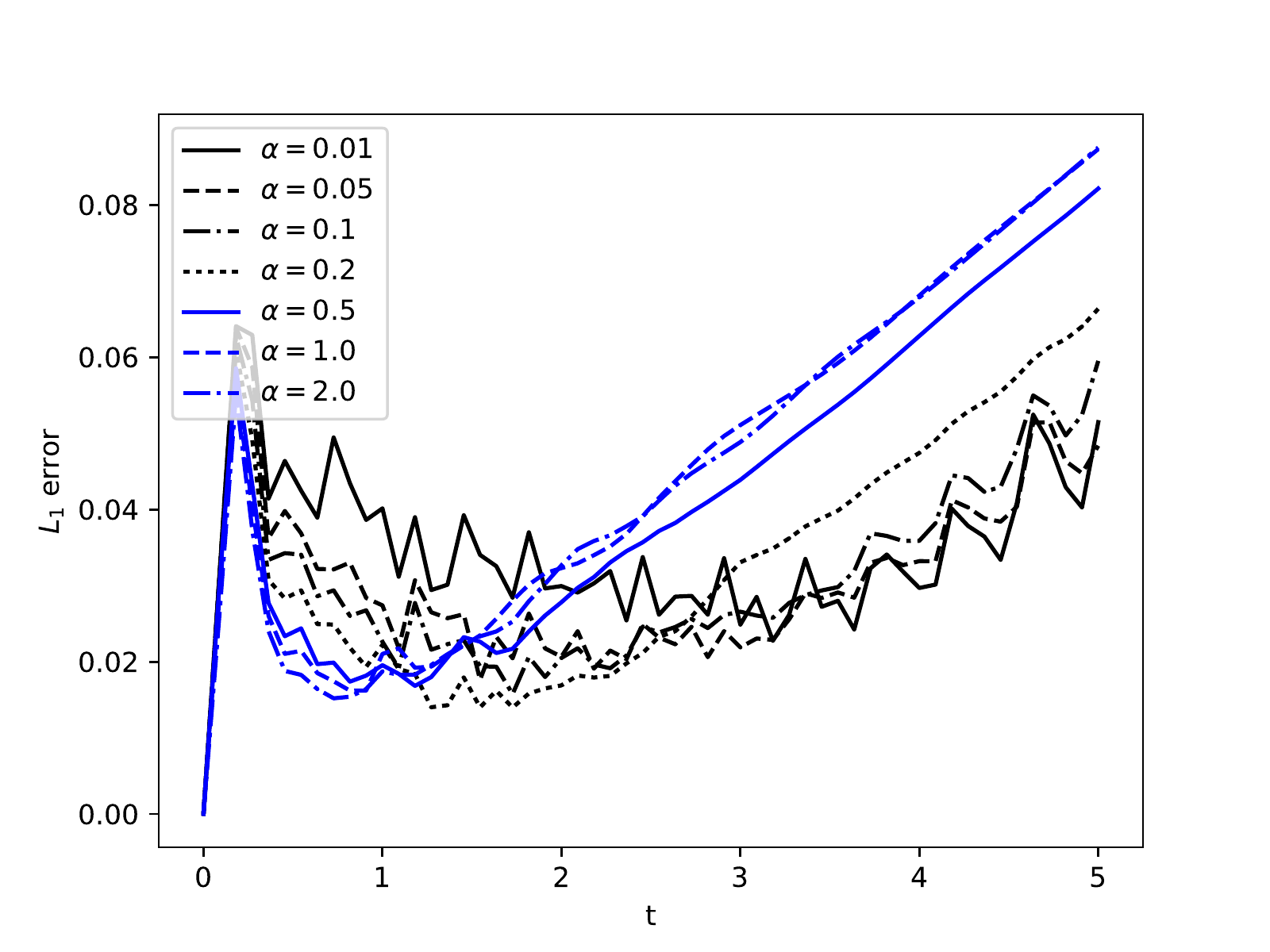}
\caption{The $L_1$ error of the velocity magnitude versus $t$ for
  different choices of $\alpha$ with $Re=100, n_x=25$ while using the
  EDAC scheme.}
\label{fig:tgv:re100_alpha}
\end{figure}

\begin{figure}
\centering
\includegraphics[width=10cm]{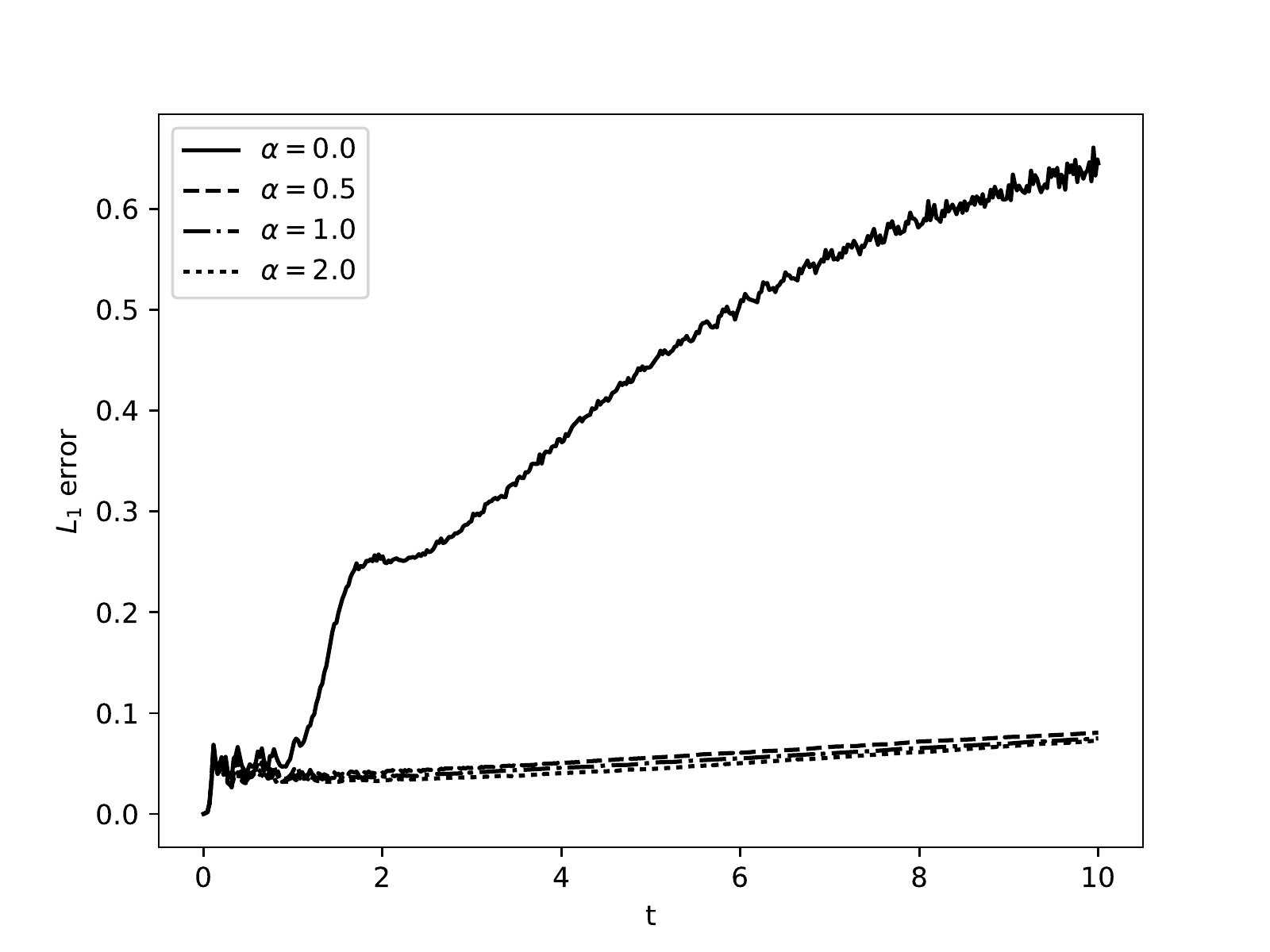}
\caption{The $L_1$ error of the velocity magnitude versus $t$ for different
  choices of $\alpha$ at $Re=10000$ while using the EDAC scheme.  When
  $\alpha=0$ is used the $\nu_{edac}$ is set to the fluid viscosity $\nu$.}
\label{fig:tgv:re10k_alpha}
\end{figure}

In order to study the sensitivity of the simulations to variations in the
parameter $\alpha$ (equation~\ref{eq:edac-nu}) used for the diffusion of the
pressure in the EDAC scheme, a few simulations with $n_x=25, Re=100$ for
different values of $\alpha$ are performed. The results are shown in
Fig.~\ref{fig:tgv:re100_alpha}. For the variation of alpha by two orders of
magnitude, the variation in the $L_1$ error is of the order of 0.04 which is
less than the order of the error produced by the TVF scheme with 16 times as
many particles. We perform a similar study with $Re=10000$ with a higher
particle resolution and present the results in Fig.~\ref{fig:tgv:re10k_alpha}.
From both these cases we are able to say that using an $\alpha$ of either 0.5
or 1.0 is reasonable.

\begin{figure}
\centering
\includegraphics[width=10cm]{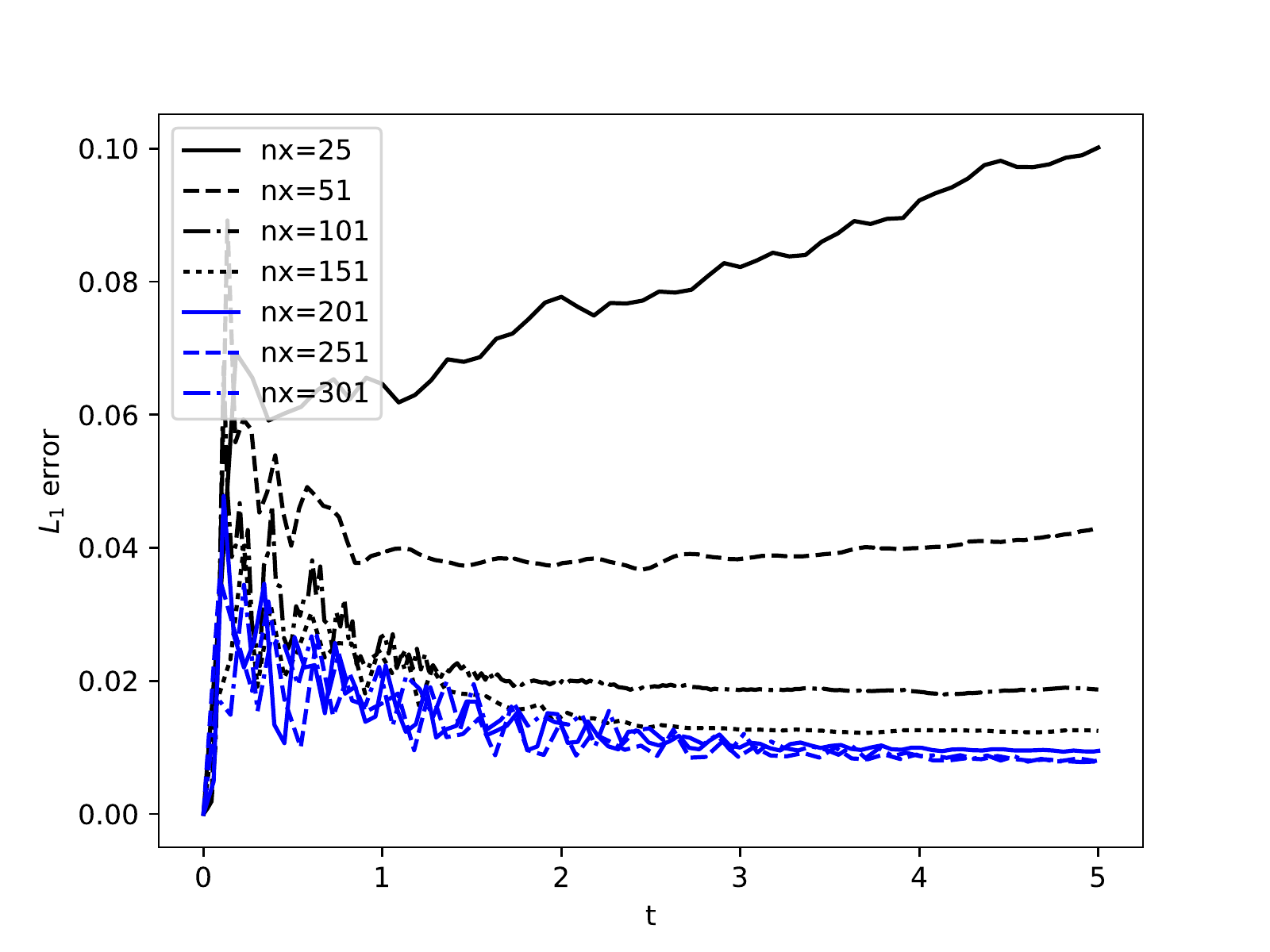}
\caption{The $L_1$ error of the velocity magnitude versus $t$ for
  different choices of $n_x$ at $Re=1000$ while using the EDAC
  scheme.}
\label{fig:tgv:re1000_conv}
\end{figure}

\begin{figure}
\centering
\includegraphics[width=10cm]{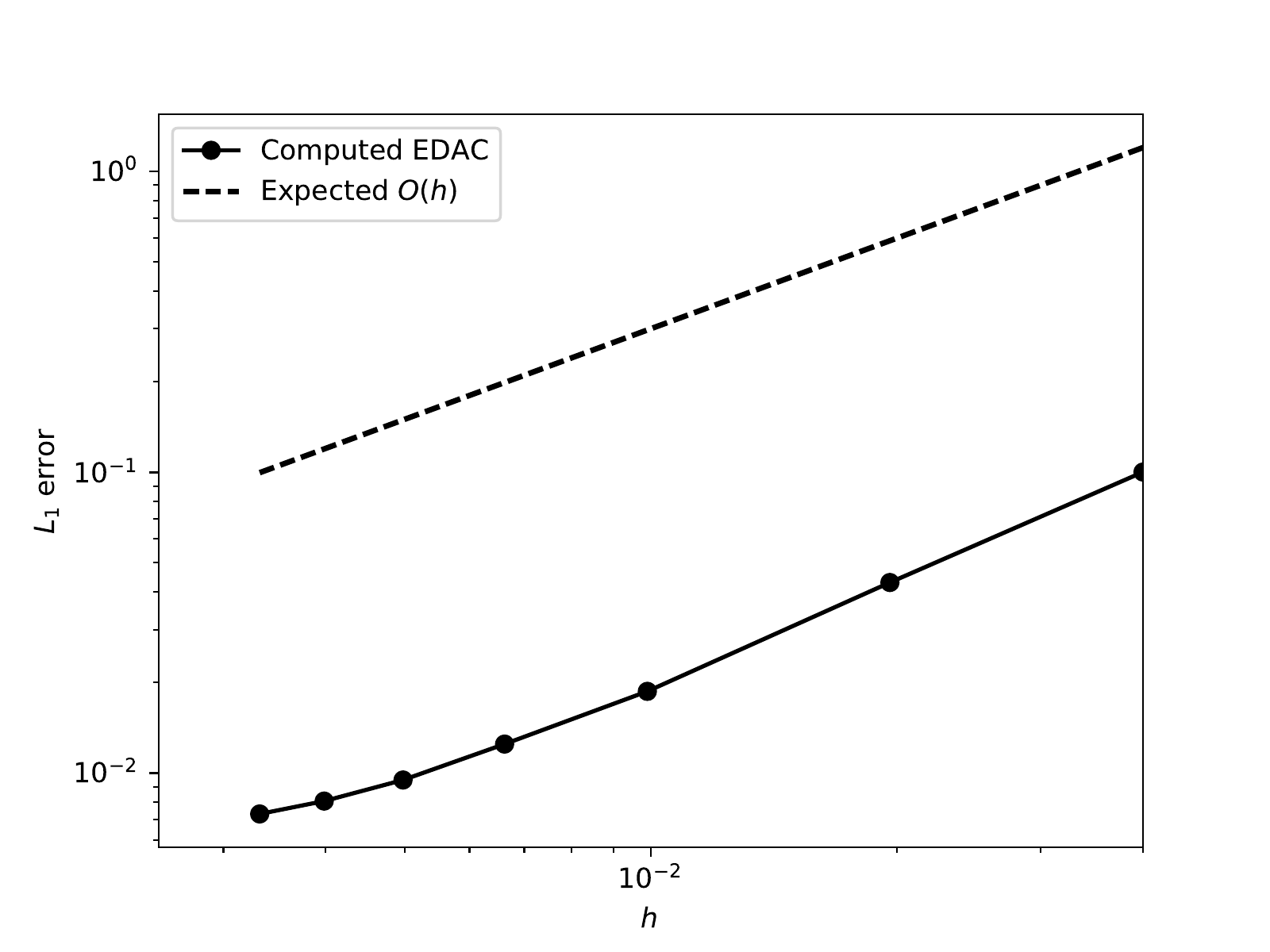}
\caption{The $L_1$ error of the velocity magnitude at $t=5$ versus $h$
  at $Re=1000$ for the EDAC scheme.  The dashed line shows the convergence of an
ideal scheme with first order convergence.}
\label{fig:tgv:re1000_conv_loglog}
\end{figure}

Fig.~\ref{fig:tgv:re1000_conv} shows a convergence study for this problem with
$Re = 1000$ and $\alpha = 1.0$. The particle spacing is increased from $n_x =
25$ to $n_x = 301$. Convergence in the $L_1$ norm for the velocity magnitude
is seen in Fig.~\ref{fig:tgv:re1000_conv_loglog}. The scheme appears to
demonstrate first order convergence. However, it can be seen that when the
$n_x$ is more than 200 the convergence rate drops. When a standard WCSPH
scheme is used with a continuity equation the behavior is similar in that the
convergence rate drops as the number of particles is increased. It is
known~\cite{sph:2d-turbulence:robinson:ijnmf2002,sph-convergence:zhu:TAJ:2015}
that as the number of particles increase one must also increase the parameter
$h$ such that the number of neighbors increase in order to have convergence.
The results suggest that the new scheme is accurate, albeit suffering from the
typical convergence related issues with traditional SPH schemes.

\begin{figure}
\centering
\includegraphics[width=10cm]{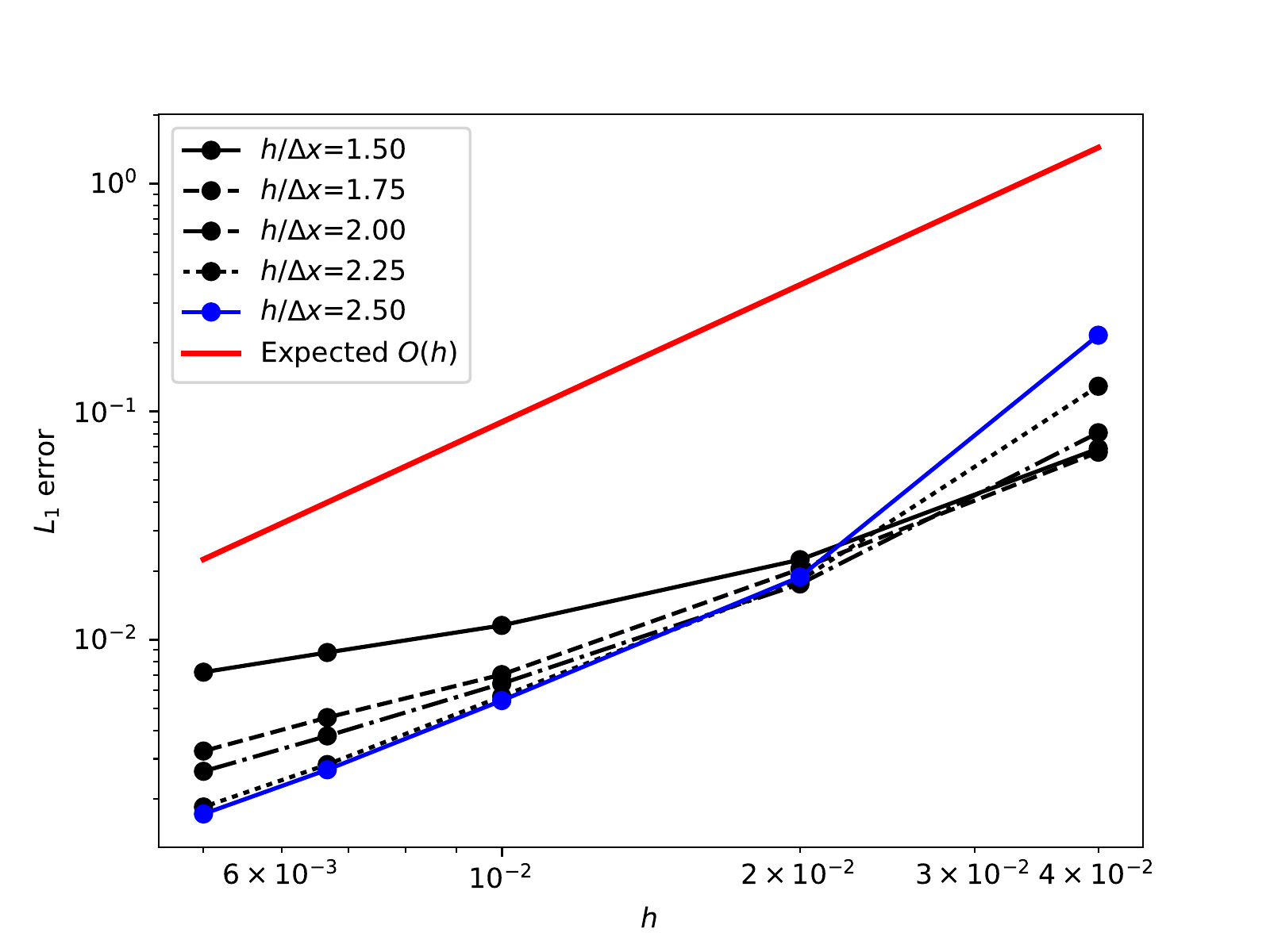}
\caption{The $L_1$ error of the velocity magnitude versus $h$ at $t=2.5$ for
  different choices of $h/\Delta x$ at $Re=1000$ while using the EDAC scheme.}
\label{fig:tgv:re1000_h_dx_conv}
\end{figure}

In order to study the error variation as the parameter $h/\Delta x$ is
changed, simulations are made at a $Re=1000$ with varying $n_x$ values. For
this case, the Wendland Quintic (C2) kernel is used as it is best suited when
increasing $h/\Delta x$ (see \cite{dehnen-aly-paring-instability-mnras-2012}).
Fig.~\ref{fig:tgv:re1000_h_dx_conv} plots the results obtained. The result
clearly shows that as $h/\Delta x$ is increased, the accuracy increases and
the convergence rate can be maintained.

\begin{figure}
\centering
\includegraphics[width=10cm]{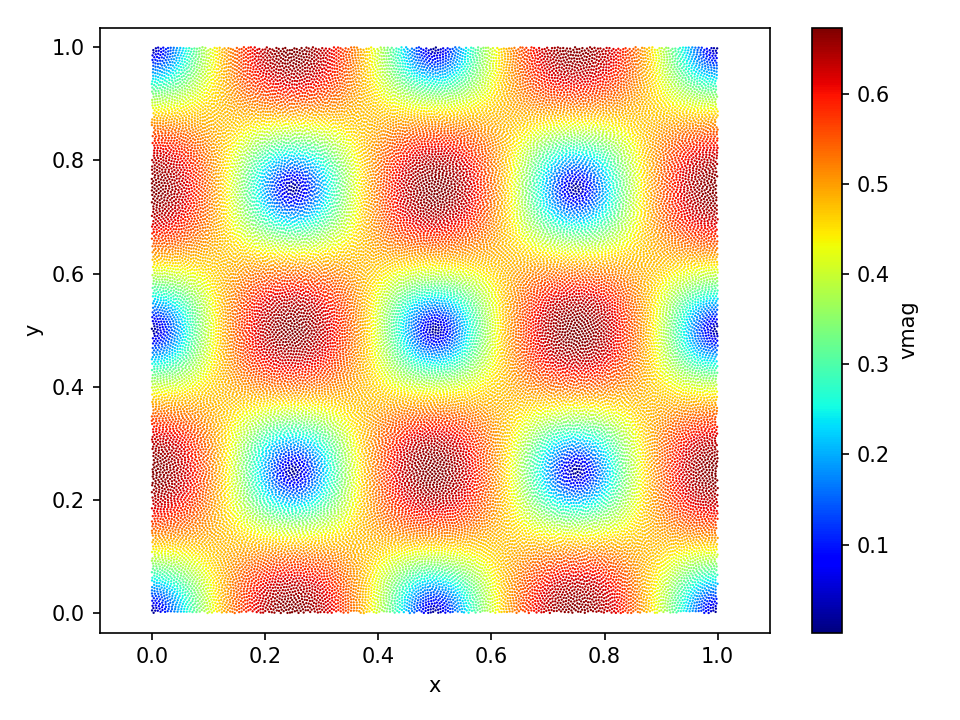}
\caption{The distribution of particles at $t=5$ for the simulation using the
  EDAC scheme with $n_x=201$ at $Re=1000$. The colors indicate the velocity
  magnitude.}
\label{fig:tgv:edac-nx201-vmag}
\end{figure}

In Fig.~\ref{fig:tgv:edac-nx201-vmag} we plot the distribution of particles at
the end of 5 seconds for the simulation at $Re=1000$ with $n_x=201$. This
shows that the distribution of particles even at a very high resolution does
not have any visible particle clustering.

\begin{figure}
\centering
\includegraphics[width=10cm]{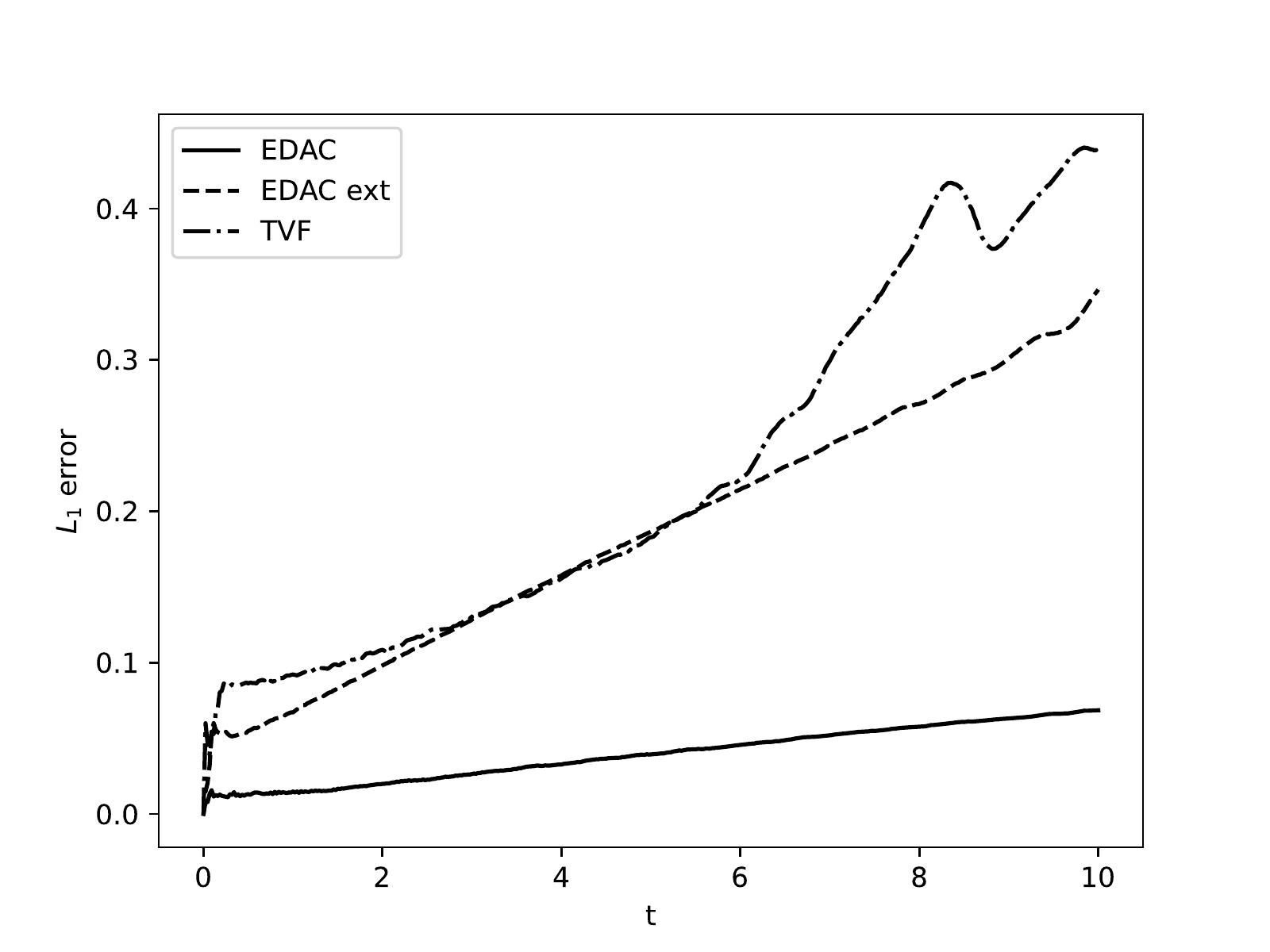}
\caption{The $L_1$ error of the velocity magnitude versus $t$ at $Re=10000$
  for the different schemes, TVF, EDAC, and EDAC external.}
\label{fig:tgv:re10k_comp}
\end{figure}

It is useful to compare the performance of the proposed scheme at high
Reynolds numbers. To this end, simulations are performed at $Re=10000$. A
100x100 grid of particles is used and with a small random initial perturbation
to the particles (the maximum perturbation of $\Delta x/5$ is chosen). The
TVF, EDAC external and EDAC TVF schemes are compared. As can be seen in
Fig.~\ref{fig:tgv:re10k_comp}, the new EDAC schemes perform very well. The
EDAC TVF scheme (labeled as EDAC) significantly outperforms the TVF scheme.
The standard EDAC scheme (Section~\ref{sec:edac-ext}) performs slightly better
than the TVF.

In Fig.~\ref{fig:tgv:re10k_alpha} the Reynolds number is set to 10000 with
$n_x=101$ and $\alpha$ is varied. When $\alpha=0$, the physical viscosity is
used. Clearly, much better results are produced when the suggested numerical
viscosity value is used instead of the physical viscosity. When the suggested
value is used the results are not too sensitive to changes in $\alpha$ around
the value of $1$.

The results show that the new scheme works well and outperforms the TVF. They
justify the use of the numerical viscosity, equation~\eqref{eq:edac-nu},
instead of the physical viscosity while diffusing the pressure. It is also
important to note that unlike the TVF, the EDAC scheme works just as well when
no initial random perturbation is given to the particles.

To provide some idea of the time taken for the simulation with different
schemes we consider the case where $n_x=50$, and $Re=100$. Running the
simulation up to $t=5s$ with the WCSPH took around 129 seconds, the TVF took
137 seconds, the standard EDAC took 151 seconds, and EDAC TVF formulation took
169 seconds. These simulations were made on a quad-core Intel i7-4770 CPU
running at 3.4 GHz. The new scheme is about 25\% slower than the standard
WCSPH scheme which is reasonable considering the significant improvement in
the results. However, the new scheme has not been optimized for performance
and the above times are to illustrate the general performance.

\subsection{Lid-Driven-Cavity}
\label{sec:ldc}

The next test problem considered is the classical Lid-Driven-Cavity (LDC)
problem, which can be a fairly challenging problem to simulate with SPH. The
setup is simple, a unit square box with no-slip walls on the bottom, left and
right boundaries. The top wall is assumed to be moving with a uniform
velocity, $V_{\text{lid}}$, which sets the Reynolds number for the problem
($Re = \frac{V_{\text{lid}}}{\nu}$). The present scheme is studied for two
Reynold's numbers $Re=100$, and $Re=1000$ and the results are compared to
those of Ghia et al.~\cite{ldc:ghia-1982}.

The quintic spline kernel is used with $h=\Delta x$. The PEC type
predictor-corrector integrator is used with a fixed time-step, chosen
according to equation~\eqref{eq:adaptive_timestep}. In addition, $\alpha =
0.5$ for all the SPH simulations. Since this problem does not involve
free-surfaces, the TVF-EDAC scheme can be used (Section~\ref{sec:edac-int}).

The discretization in terms of the number of particles is dependent on the
Reynold's number. A uniform distribution of particles ($\Delta x = \Delta y$)
is used, with the maximum resolution of $50 \times 50$, and $100 \times 100$
for the $Re = 100$, and $Re = 1000$ cases respectively. The timesteps are
chosen according to equation~\eqref{eq:adaptive_timestep} as before. For each
case, the code is run for a sufficiently long time to reach a steady state.
This was found by looking at the kinetic energy of the entire fluid and
checking if it is essentially constant. The velocity plots are made by
averaging over the last $10$ saved time-step results. The data is saved every
$500$ time-steps.

\begin{figure}[h]
\centering
\includegraphics[width=14cm]{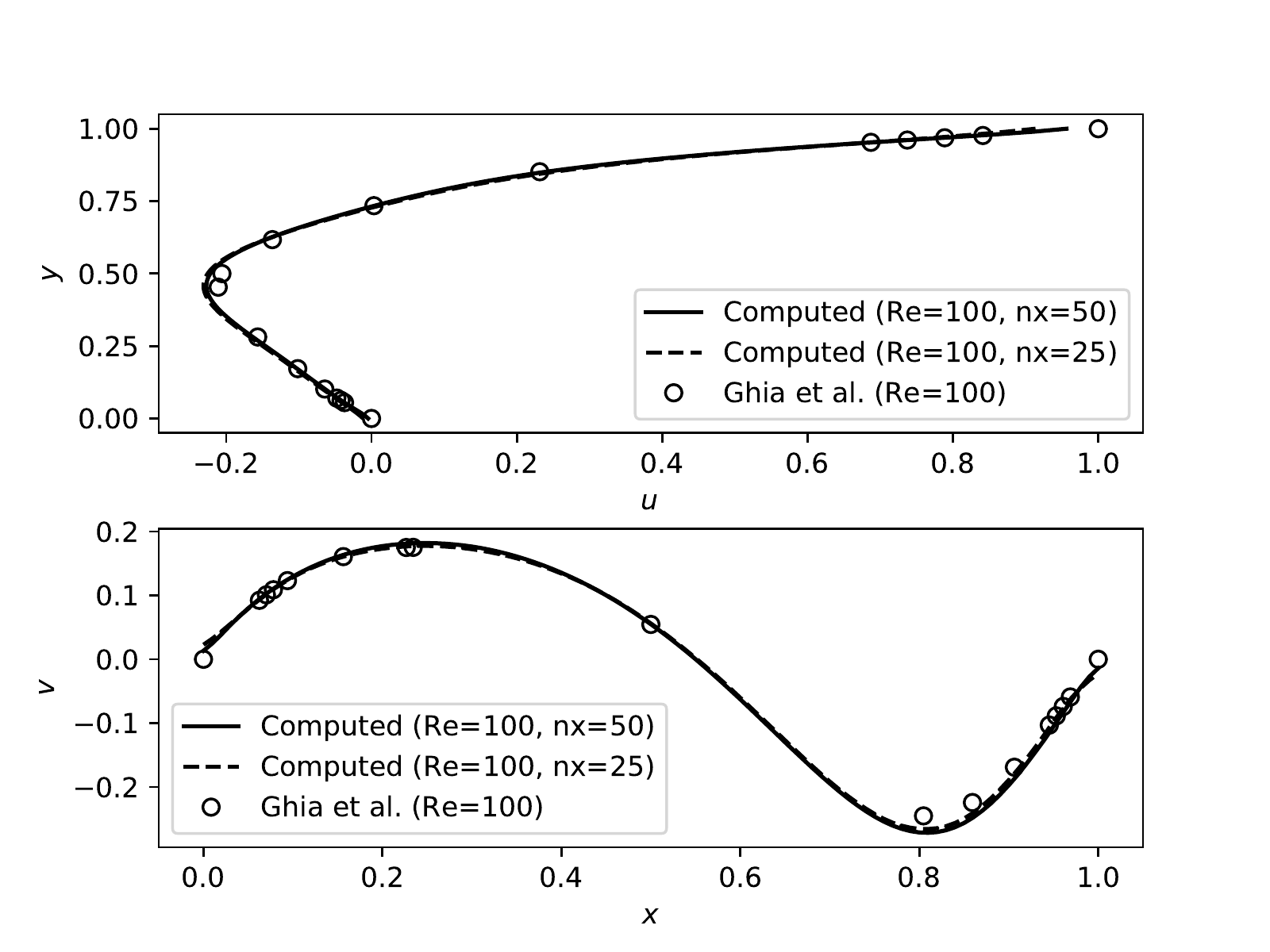}
\caption{The velocity profiles $u$ vs.\ $y$ and $v$ vs.\ $x$ for the
  lid-driven-cavity problem at $Re=100$. The results are compared with those
  of \citet{ldc:ghia-1982}.}
\label{fig:ldc:uv_re100}
\end{figure}

\begin{figure}[h]
\centering
\includegraphics[width=14cm]{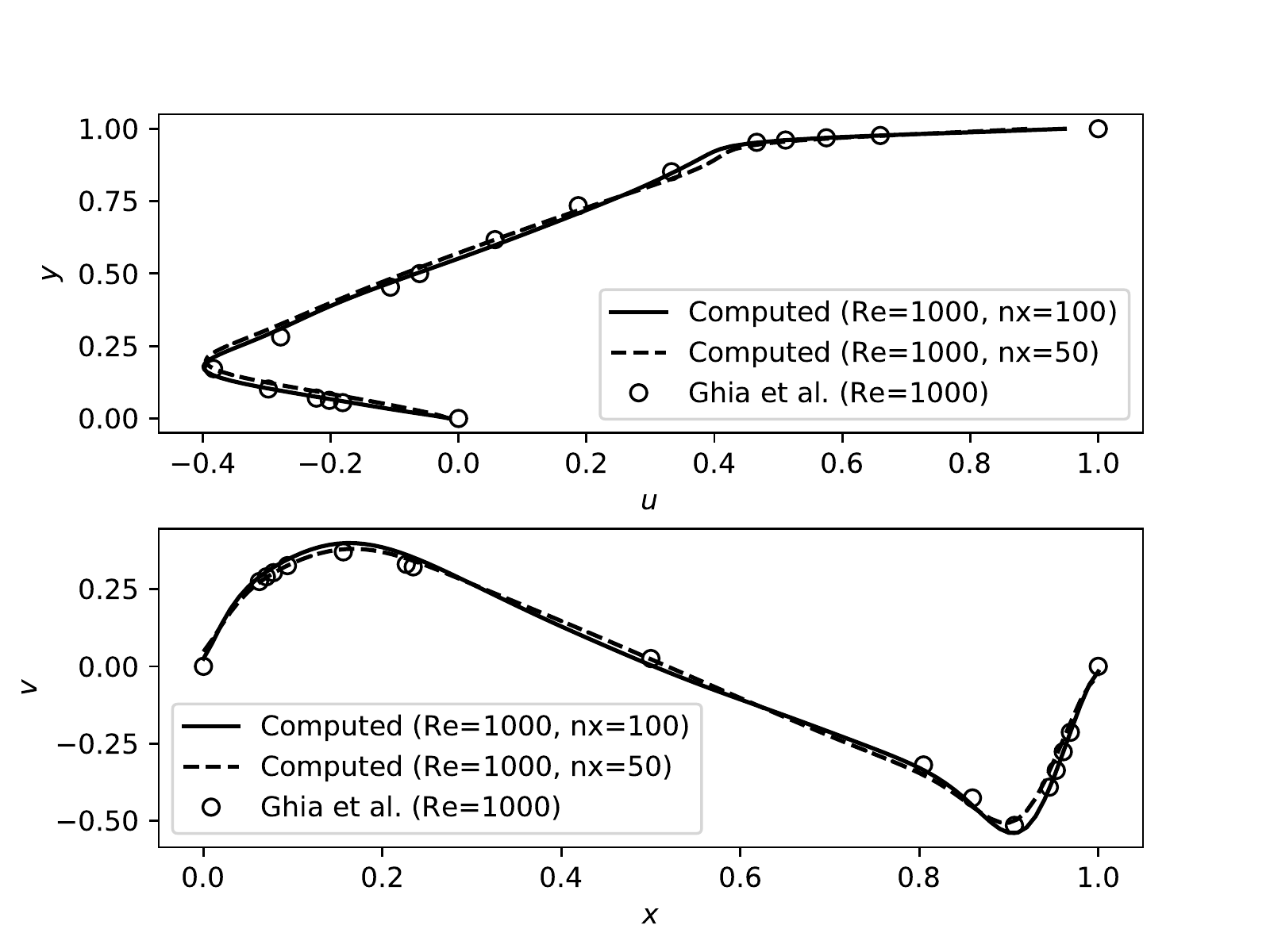}
\caption{The velocity profiles $u$ vs.\ $y$ and $v$ vs.\ $x$ for the
  lid-driven-cavity problem at $Re=1000$. The results are compared with those
  of \citet{ldc:ghia-1982}.}
\label{fig:ldc:uv_re1000}
\end{figure}

Fig.~\ref{fig:ldc:uv_re100} shows the results for two different resolutions at
$Re=100$ and Fig.~\ref{fig:ldc:uv_re1000} shows those for $Re=1000$. The
results are in good agreement with those of Ghia et al.~\cite{ldc:ghia-1982}.

\subsection{Periodic lattice of cylinders}
\label{sec:lattice}

The next problem considered is a benchmark problem in a periodic square domain
with a cylinder. The periodicity implies that the fluid effectively sees a
periodic lattice of cylinders. This test was used to evaluate the TVF scheme
in~\cite{Adami2013} and identical parameters are used for the numerical
set-up. The length of the square domain is $L = 0.1m$ and the Reynold's number
is set to one. A body force, $g_x=1.5 \times 10^{-7}m/s^2$ drives the flow
along the $x$ direction. The cylinder is placed in the center of the domain
with a radius $R = 0.02m$.   The domain is periodic along both the
coordination directions. A uniform discretization is used with $100 \times
100$ particles and a quintic spline kernel with $h = \Delta x$ is used. The
PEC type predictor-corrector integrator is used with a fixed time-step chosen
using equation~\eqref{eq:adaptive_timestep}.

\begin{figure}[htpb]
\centering
\includegraphics[width=10cm]{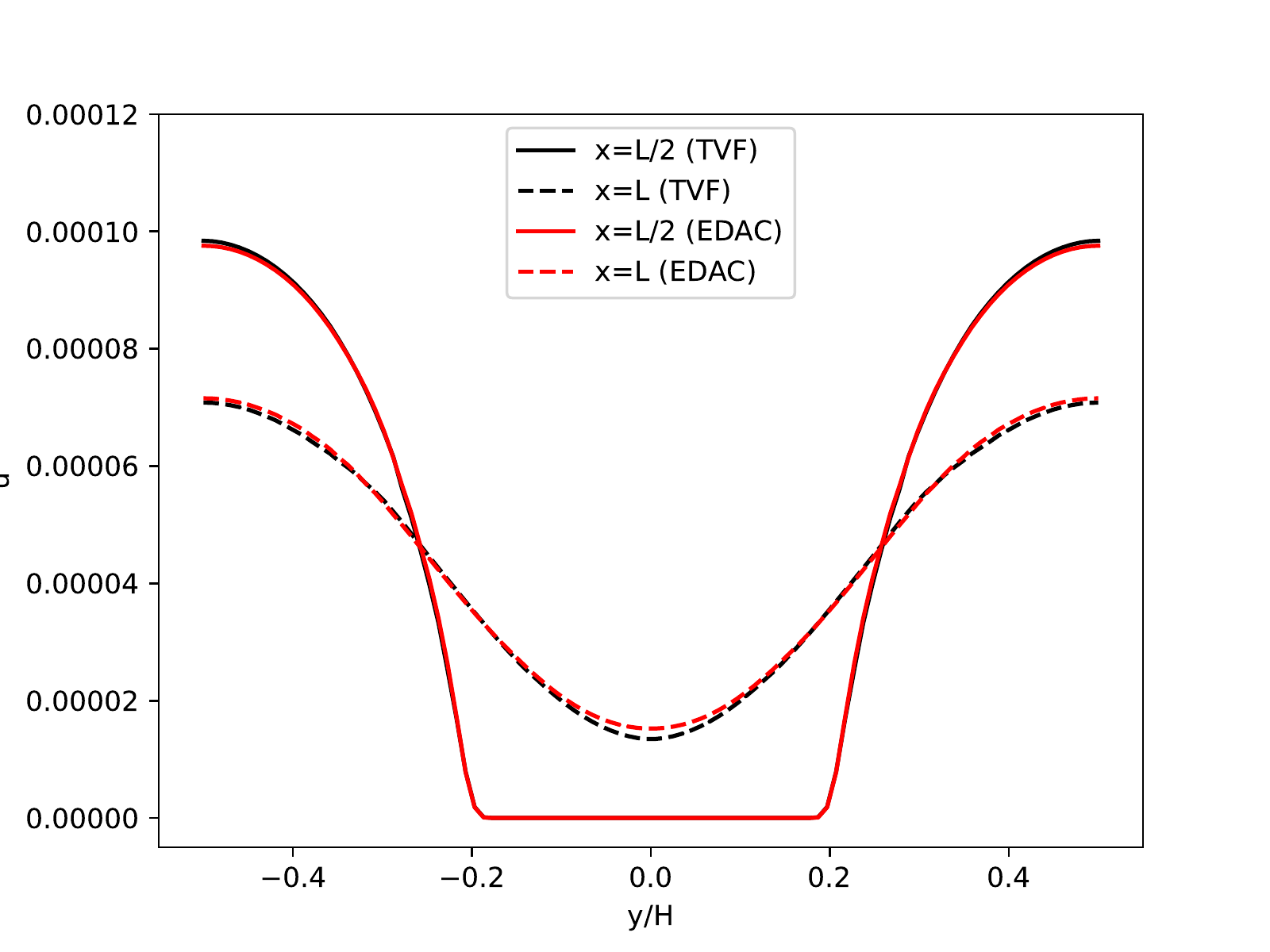}
\caption{Axial velocity profile ($u$) along the transverse ($y$) direction at
  $x=L/2$ and $x=L$ for the periodic lattice of cylinders using the EDAC (red)
  and TVF (black) schemes.}
\label{fig:lattice:u}
\end{figure}

Fig.~\ref{fig:lattice:u} shows the axial velocity profile ($u$) along the
lines $x=L/2$ and $x=L$, when using the TVF-EDAC scheme and compare the
results with the TVF scheme. The axial velocity is obtained by performing a
Shepard interpolation of the fluid particle properties on points along the
axial line. It is found that the results of the new scheme are in good
agreement with that of the TVF scheme. For this problem the TVF simulation
took around 198 seconds and the EDAC simulation took around 259 seconds.

\subsection{Periodic array of cylinders}
\label{sec:periodic}

\begin{figure}[htpb]
\centering
\includegraphics[width=10cm]{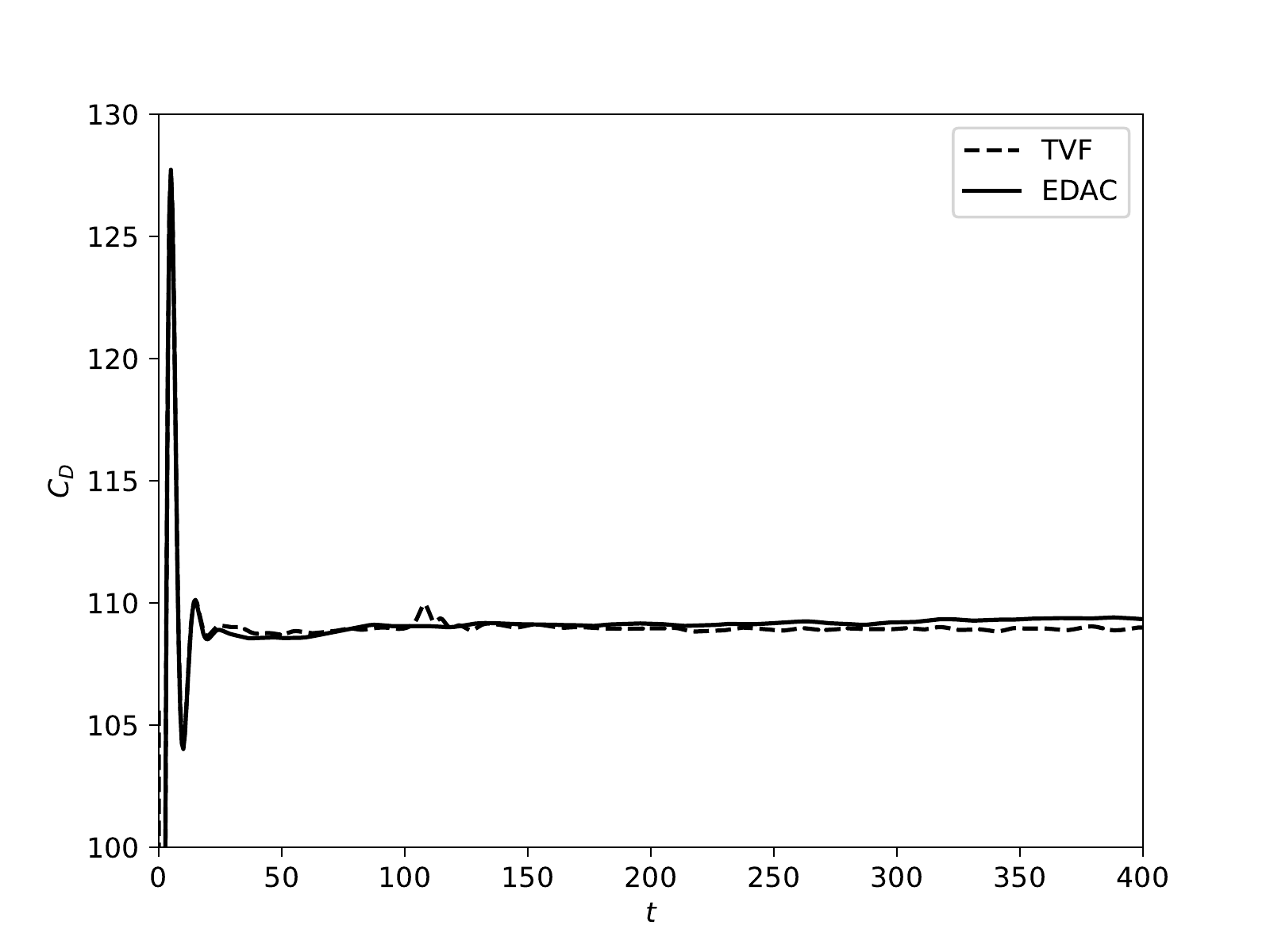}
\caption{The drag variation $C_D$ versus time for a periodic array of
  cylinders in a channel. The results from the TVF (dash) are compared with
  those produced by the EDAC (solid) scheme.}
\label{fig:periodic_cyl:cd}
\end{figure}

\begin{figure}[htpb]
\centering
\includegraphics[width=10cm]{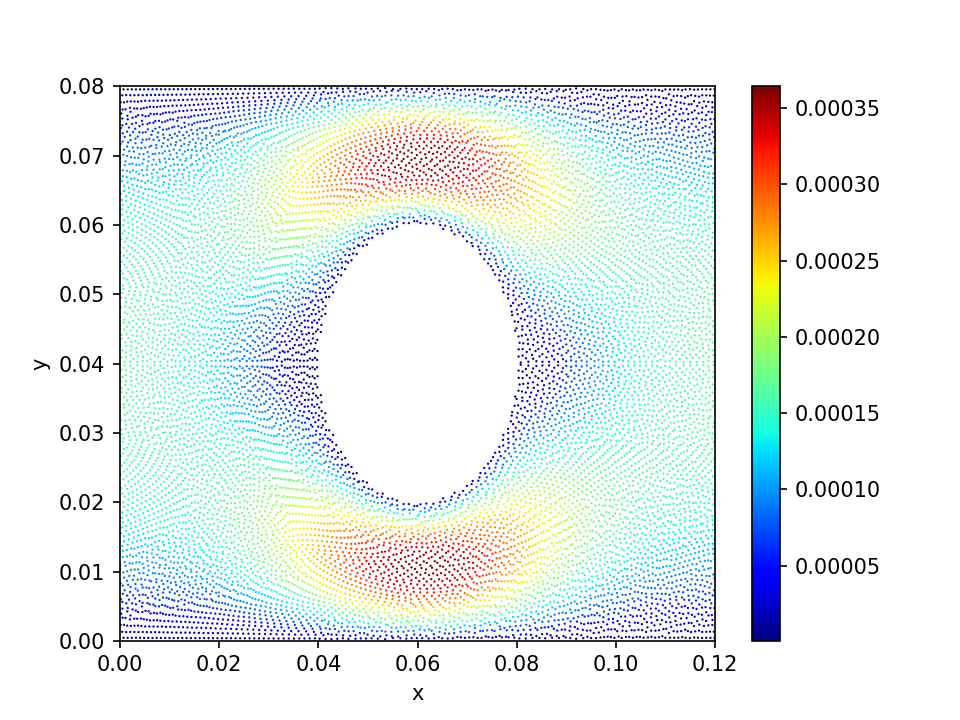}
\caption{The distribution of particles at the final time produced by the EDAC
  scheme.  The color indicates the velocity magnitude.}
\label{fig:periodic_cyl:particles}
\end{figure}

The next benchmark is similar to the periodic lattice of cylinders but with
no-slip wall boundary conditions along the top and bottom walls. The domain is
periodic in the $x$ direction, driven by a body force $g_x= 2.5\times
10^{-4}m/s^2$. A rigid cylinder with radius $R = 0.2m$ is placed in the center
of the channel. The length of the channel is $L = 0.12m$ and the height is $H
= 4R$. The numerical set-up is identical to that of \citet{Adami2013} with
$n_x=144$ but with $h=1.2 \Delta x$ chosen for both schemes.
Fig.~\ref{fig:periodic_cyl:cd} shows the drag coefficient on the cylinder
generated by the TVF and the new scheme. Fig.~\ref{fig:periodic_cyl:particles}
shows the distribution of the particles at the final time produced by the EDAC
scheme. The particles are homogeneously distributed as would be expected. The
particle distribution is very similar to that produced by \citet{Adami2013}.

Note that for this problem, when using $c=0.1\sqrt{g_xR}$, as recommended by
\citet{Adami2013}, the particle positions diverge when using the TVF
formulation. Instead, in order to reproduce the results of \citet{Adami2013}
the value is set to $c= 0.02 m/s$ as recommended by
\citet{sph:ellero-adams:ijnmf:2011}.

The present results suggest that the EDAC scheme performs well for all of the
internal flow cases. A few standard free-surface problems are considered next.

\subsection{Elliptical drop}
\label{sec:ed}

The elliptical drop problem is a classic problem that was first solved in the
context of SPH by Monaghan~\cite{sph:fsf:monaghan-jcp94}. The problem studies
the evolution of a circular drop of inviscid fluid having unit radius in free
space with the initial velocity field given by $-100x\hat{i} + 100 y \hat{j}$.
No specific boundary conditions are applied on the outer surface of the
elliptical drop as they are treated as a free surface. The incompressibility
constraint on the fluid enables a derivation for evolution of the semi-major
axis of the ellipse. The problem is simulated with the standard WCSPH scheme
where an artificial viscosity with $\alpha=0.1$ is used. The particle spacing
is chosen to be $\Delta x = 0.025m$. A Gaussian kernel is used for the WCSPH
with $h=1.3 \Delta x$. The value of $\gamma=7$. The speed of sound is set to
$1400 m/s$ and $\rho=1.0kg/m^3$. For the EDAC case, a quintic spline kernel is
used with $h=1.2 \Delta x$. $\alpha=0.5$ for the calculation of $\nu_{edac}$.
An Evaluate-Predict-Evaluate-Correct (EPEC) integration scheme is used for the
WCSPH scheme whereas a Predict-Evaluate-Correct (PEC) integrator is used for
the new scheme and the results are compared.

\begin{figure}
\centering
\includegraphics[width=10cm]{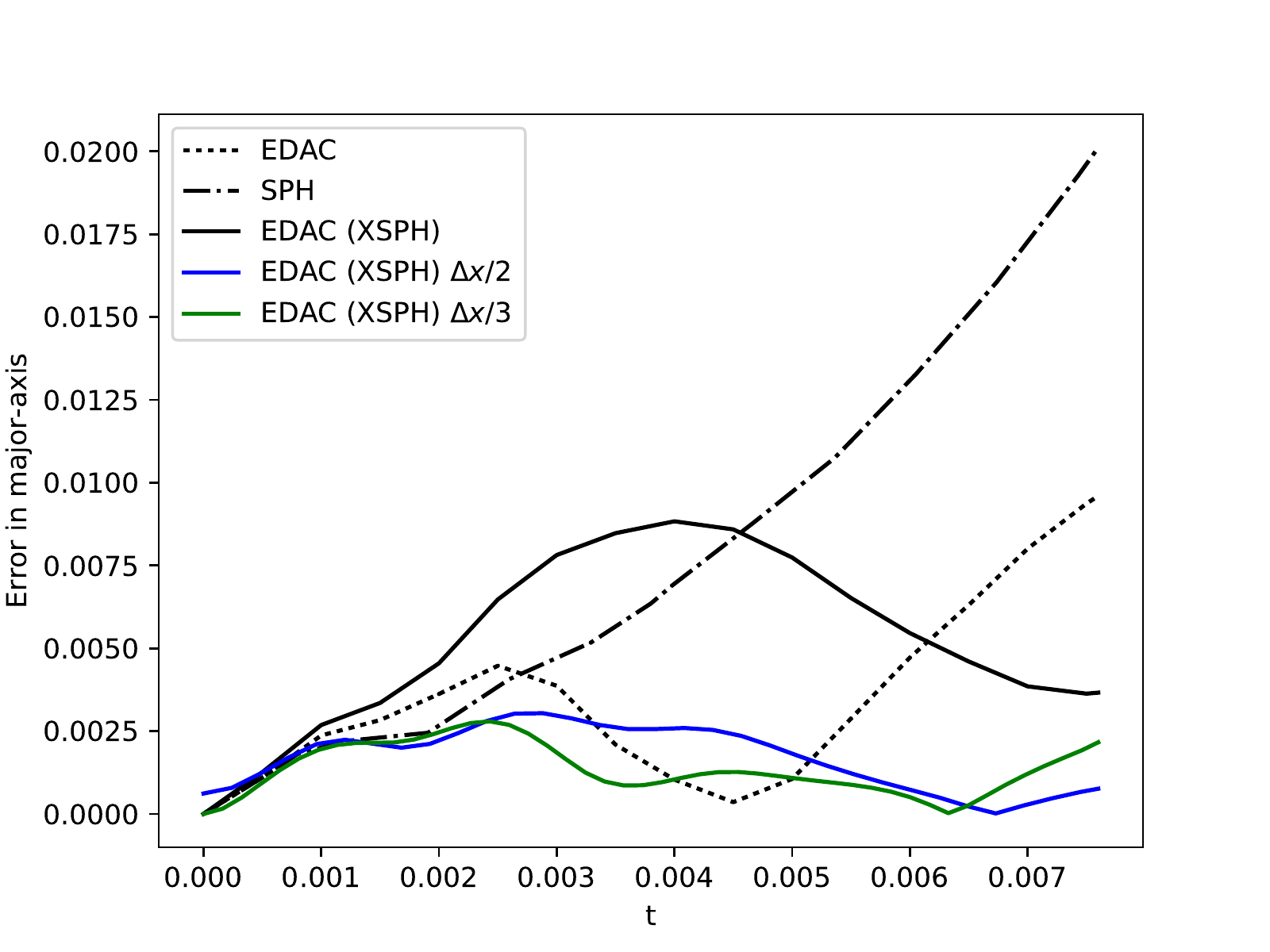}
\caption{The error in the computed size of the semi-major axis
  compared for the standard SPH, EDAC and the EDAC with the use of XSPH.}
\label{fig:ed:major-axis}
\end{figure}

In Fig.~\ref{fig:ed:major-axis}, the semi-major axis of the ellipse is
compared with the exact solution. The standard EDAC scheme
(Section~\ref{sec:edac-ext}) is used to simulate the problem. No artificial
viscosity is used for the EDAC scheme. Artificial viscosity is used for the
WCSPH implementation with a value of $\alpha=0.1, \beta=0.0$. One EDAC
simulation is performed using the XSPH
correction~\cite{sph:xsph:monaghan-jcp89} and one without it. Two additional
cases of the EDAC along with the XSPH correction with a resolution of $\Delta
x/2$ and $\Delta x/3$ are also performed. The absolute error in the size of
the semi-major axis with time is used as a metric to compare the results. As
can be seen, the EDAC scheme performs better than the standard SPH both with
and without the XSPH correction.

\begin{figure}
\centering
  \includegraphics[width=10cm]{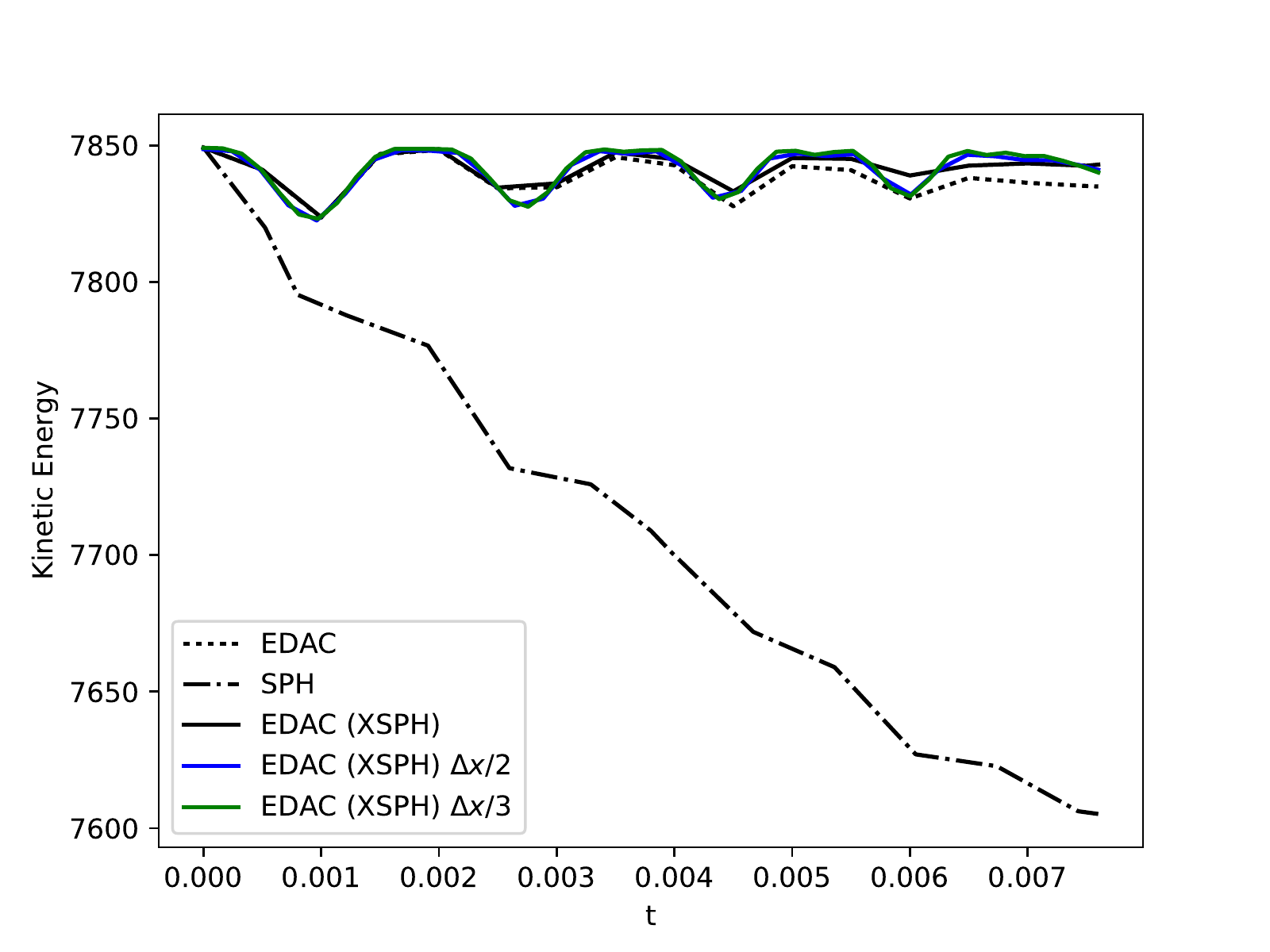}
\caption{The kinetic energy of the elliptical drop computed by
  different schemes.}
\label{fig:ed:ke}
\end{figure}

In Fig.~\ref{fig:ed:ke}, the kinetic energy of the fluid is computed and
plotted versus time. It is to be noted that one may obtain the exact kinetic
energy by integrating the initial velocity field. Given a unit density and an
initial radius of unity, this amounts to approximately 7853.98 units. The
kinetic energy of the standard SPH formulation reduces due to the artificial
viscosity. The EDAC scheme on the other hand does not display any significant
loss of kinetic energy and the value is close to the exact value.

\begin{figure}
  \centering
  \begin{subfigure}[b]{0.3\textwidth}
    \includegraphics[width=1.1\textwidth]{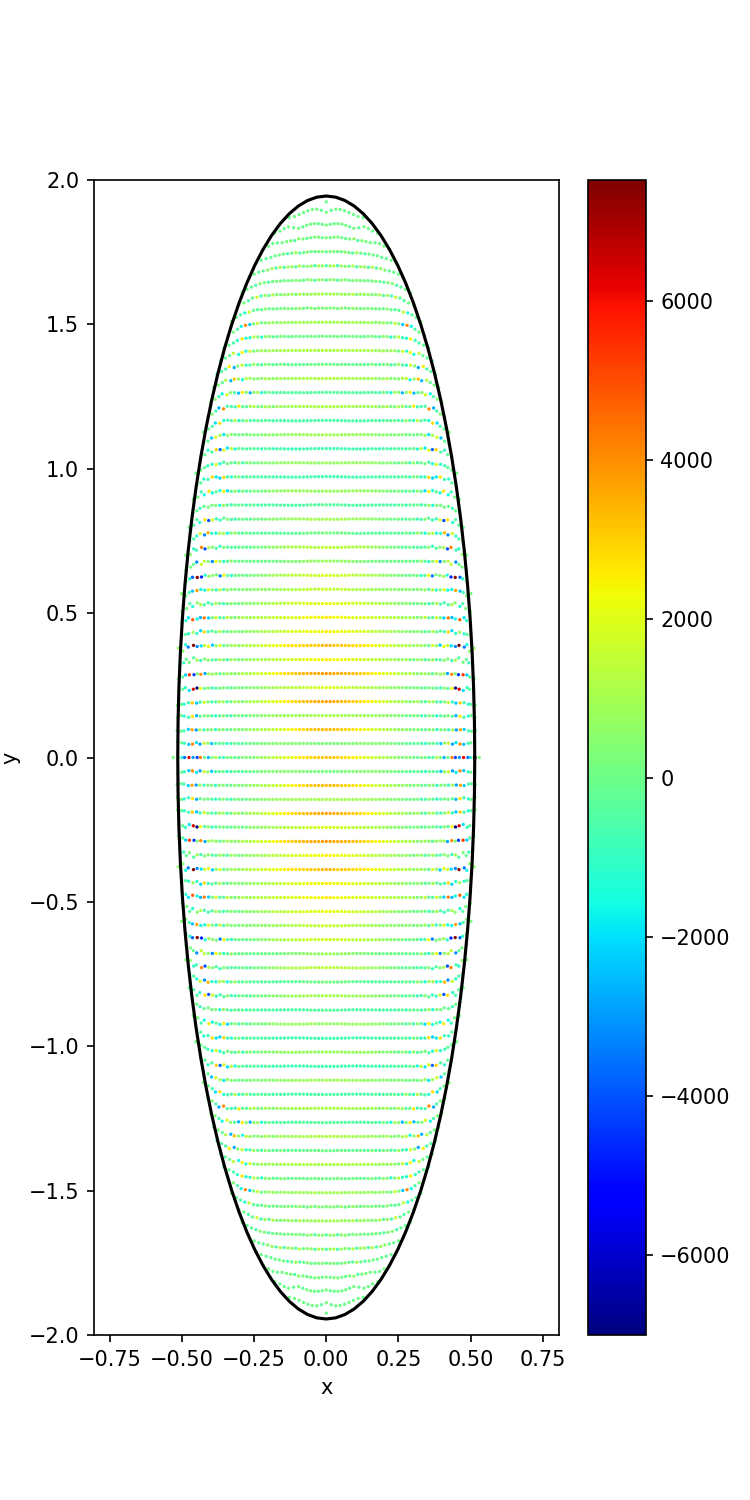}
    \caption{WCSPH.}
    \label{fig:ed:std-sph}
  \end{subfigure}
  ~
  \begin{subfigure}[b]{0.3\textwidth}
    \includegraphics[width=1.1\textwidth]{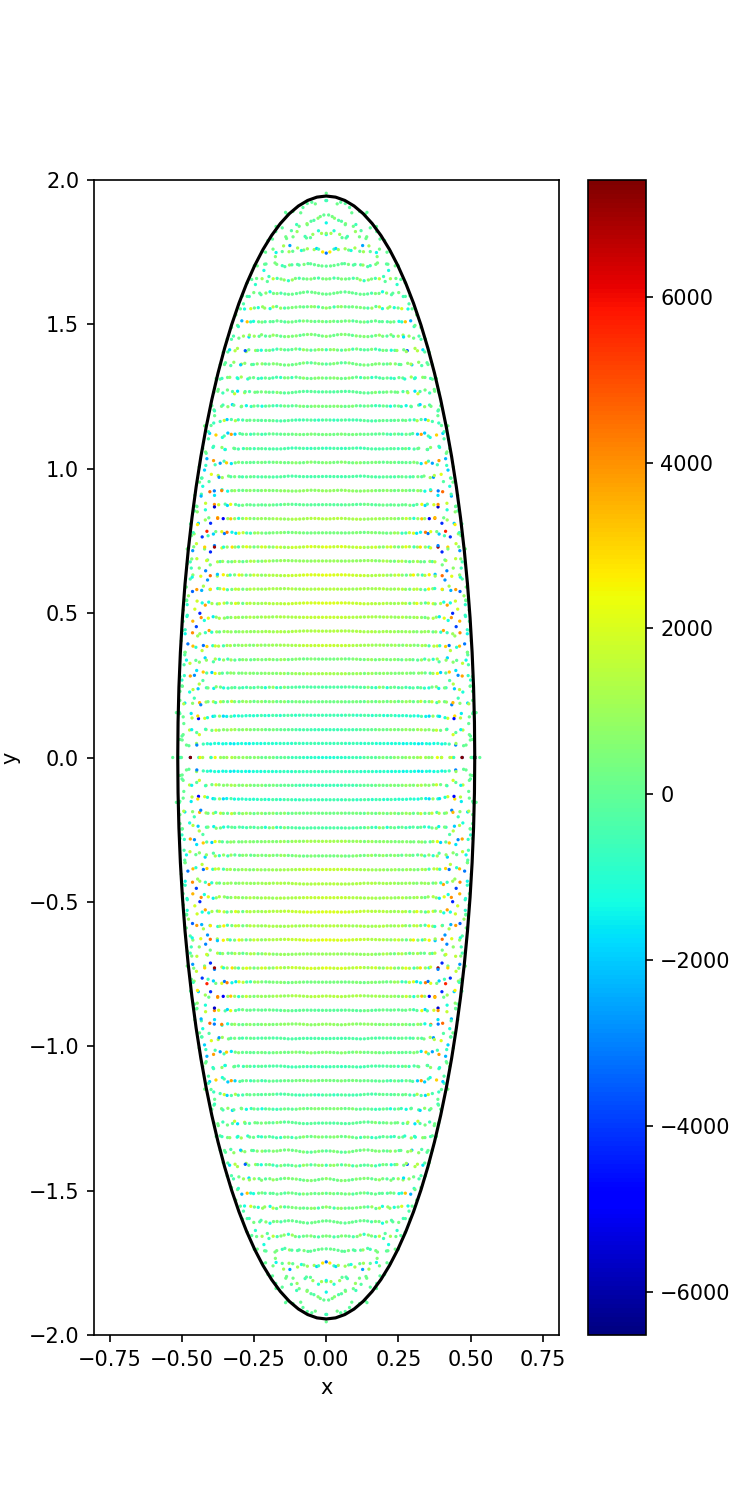}
    \caption{EDAC no XSPH.}
    \label{fig:ed:edac-particles}
  \end{subfigure}
  ~
  \begin{subfigure}[b]{0.3\textwidth}
    \includegraphics[width=1.1\textwidth]{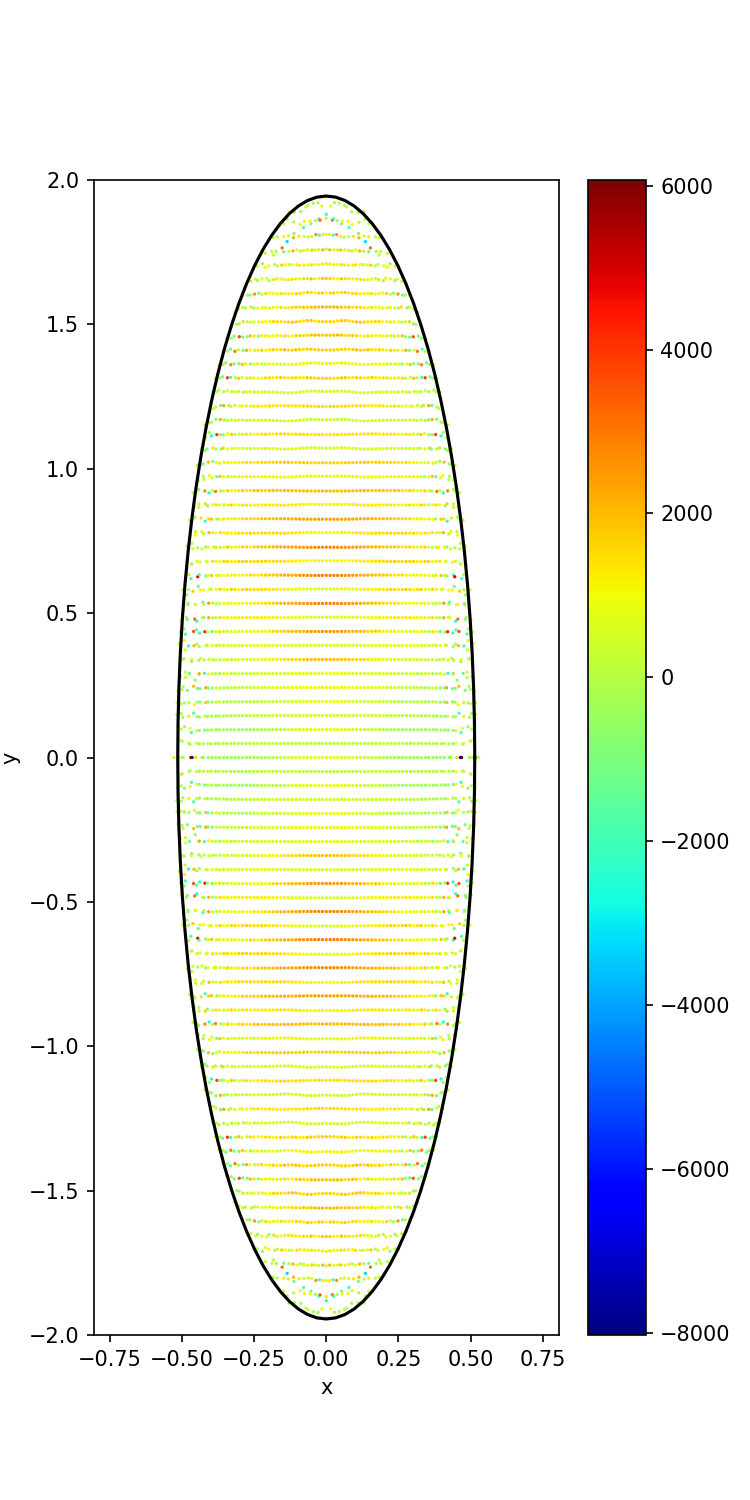}
    \caption{EDAC with XSPH.}
\label{fig:ed:edac-xsph}
  \end{subfigure}
  \caption{The distribution of particles for the elliptical drop problem at $t
    = 0.0076$ seconds. The plot (a) is with the standard WCSPH scheme with the
    use of artificial viscosity.  Plot (b) is with the EDAC without XSPH and
    (c) is EDAC with XSPH. The solid line is the exact solution and the
    colors indicated the pressure.}
  \label{fig:ed:particle-plots}
\end{figure}

Fig.~\ref{fig:ed:std-sph} plots the particle distribution as obtained by the
WCSPH simulation. The colors show the pressure distribution. The solid line is
the exact solution. Fig.~\ref{fig:ed:edac-particles} shows the same obtained
with the EDAC without the XSPH correction and Fig.~\ref{fig:ed:edac-xsph}
shows the particles and the pressure distribution using the EDAC scheme along
with the XSPH correction. The XSPH correction seems to reduce the noise in the
particle distribution. It is clear that the EDAC scheme has much lower
pressure oscillations than the WCSPH scheme even though no artificial
viscosity is used.

As can be seen, the new scheme outperforms the standard SPH scheme in general,
conserves kinetic energy, has lower pressure oscillations, and is quite robust
as there is no need for an artificial viscosity to keep the scheme stable.

\subsection{Hydrostatic tank}
\label{sec:ht}

The next example is a simple benchmark to ensure that the pressure is evolved
correctly. This benchmark consists of a tank of water held at rest with the
top of the vessel kept open as simulated by~\citet{Adami2012}. The fluid is
initialized with a zero pressure with the particles at rest. The acceleration
due to gravity is set to -1$m/s^2$, the height of the water is $0.9 m$ and the
density of the fluid is set to $1000 kg/m^3$. The maximum speed of the fluid
is taken to be $\sqrt{g H}$ and the speed of sound is set to ten times this
value. The timestep is calculated as before using these values. The
acceleration due to gravity is damped as discussed in \cite{Adami2012}. In
order to reproduce the results, the same artificial viscosity factor
$\alpha=0.24$ is used. No physical viscosity is used. The parameter $\alpha$
for the EDAC equation is set to $0.5$. The problem is simulated with the TVF
scheme (using no background pressure) as well as the EDAC scheme. No specific
boundary condition is explicitly applied on the free surface. The walls of the
tank are essentially slip-walls as the physical viscosity is zero. To compare
the results, the pressure is evaluated along a line at the center of the tank.

\begin{figure}
\centering
\includegraphics[width=10cm]{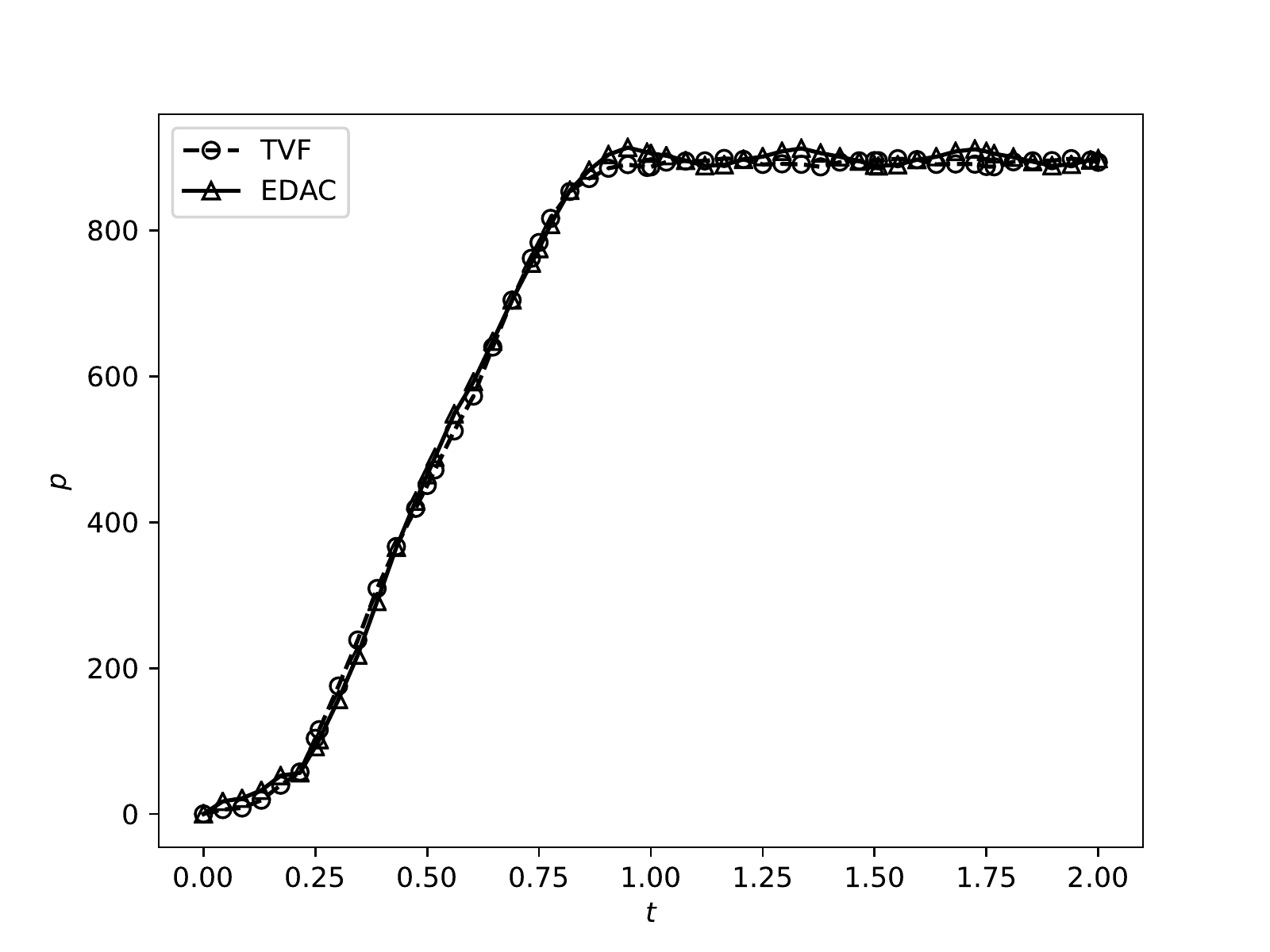}
\caption{Plot of the pressure at the bottom of the tank versus time for
  different schemes.}
\label{fig:ht:p-bottom}
\end{figure}

In Fig.~\ref{fig:ht:p-bottom} the pressure at the bottom of the tank is
plotted versus time for both the TVF scheme and the EDAC scheme. The EDAC
scheme seems to produce a bit more oscillation in the pressure but the overall
agreement is good.

\begin{figure}
\centering
\includegraphics[width=10cm]{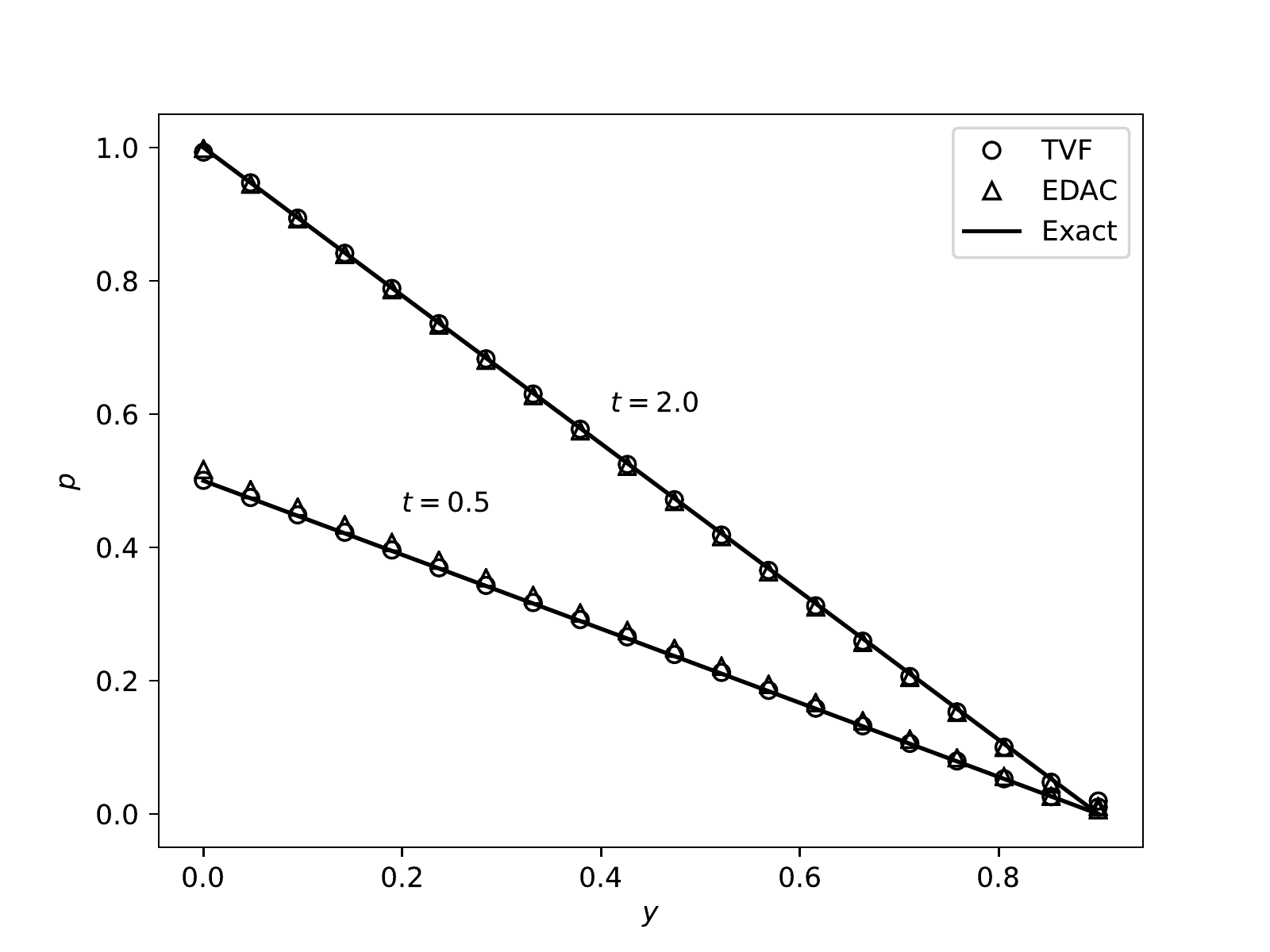}
\caption{Pressure variation with height for the different schemes at
  $t=0.5$ and $t=2.0$.}
\label{fig:ht:p-vs-y}
\end{figure}

In Fig.~\ref{fig:ht:p-vs-y}, the pressure variation with height for a line of
points at the center of the tank is plotted for different schemes at the times
$t=0.5$ and $t=2$. The agreement is very good. This shows that the EDAC scheme
produces accurate pressure distributions. In terms of execution time, the TVF
simulation takes about 24 seconds and the EDAC simulation takes about 33
seconds.

\subsection{Water impact in two-dimensions}
\label{sec:two-blocks}

\begin{figure}
  \centering
  \parbox{6cm}{\includegraphics[width=8cm]{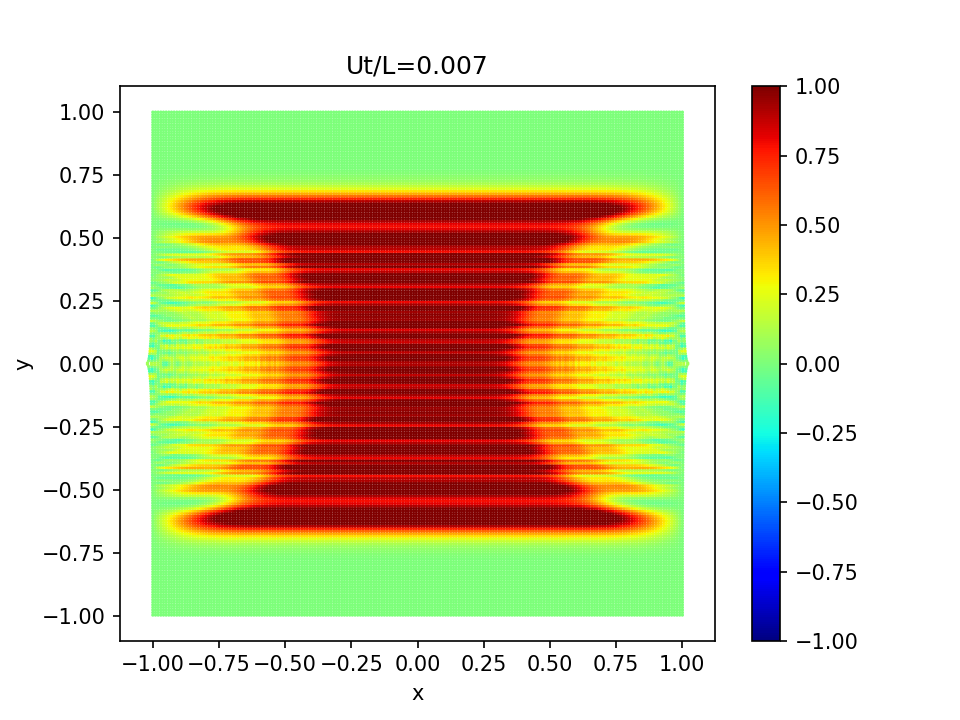}}
  \qquad
  \begin{minipage}{6cm}
    \includegraphics[width=8cm]{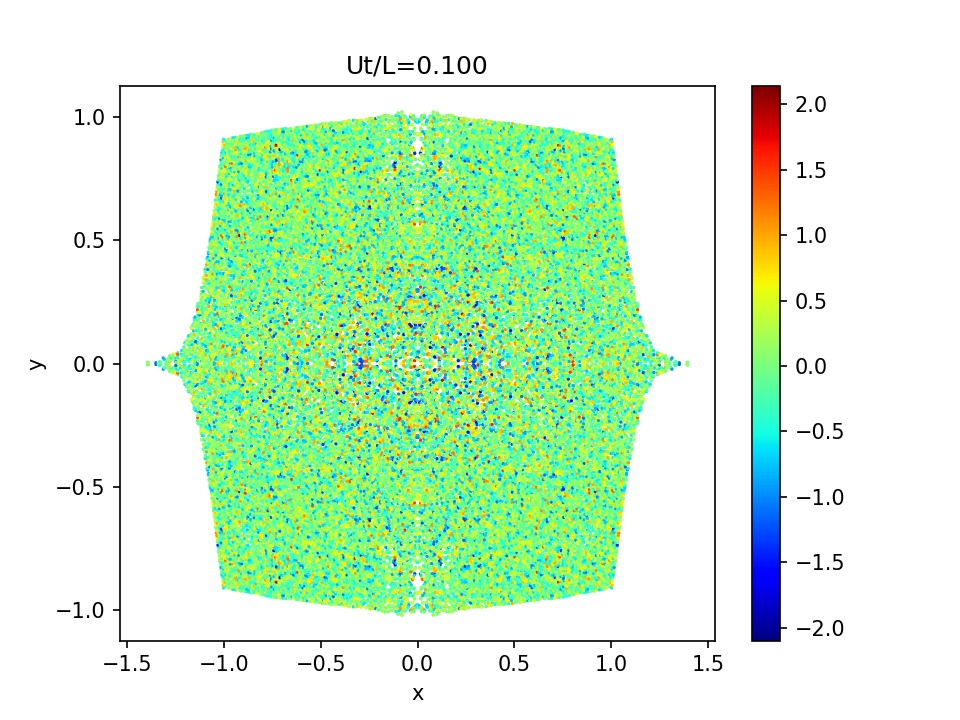}
  \end{minipage}
  \caption{Particle distribution and pressure ($p/\rho c_s U$) at $Ut/L=0.007$
    (left) and $Ut/L = 0.1$ (right) for the water impact problem with the
    standard WCSPH scheme without any artificial viscosity.}
\label{fig:two-blocks:wcsph-noavisc}
\end{figure}

\begin{figure}
  \centering
  \parbox{6cm}{\includegraphics[width=8cm]{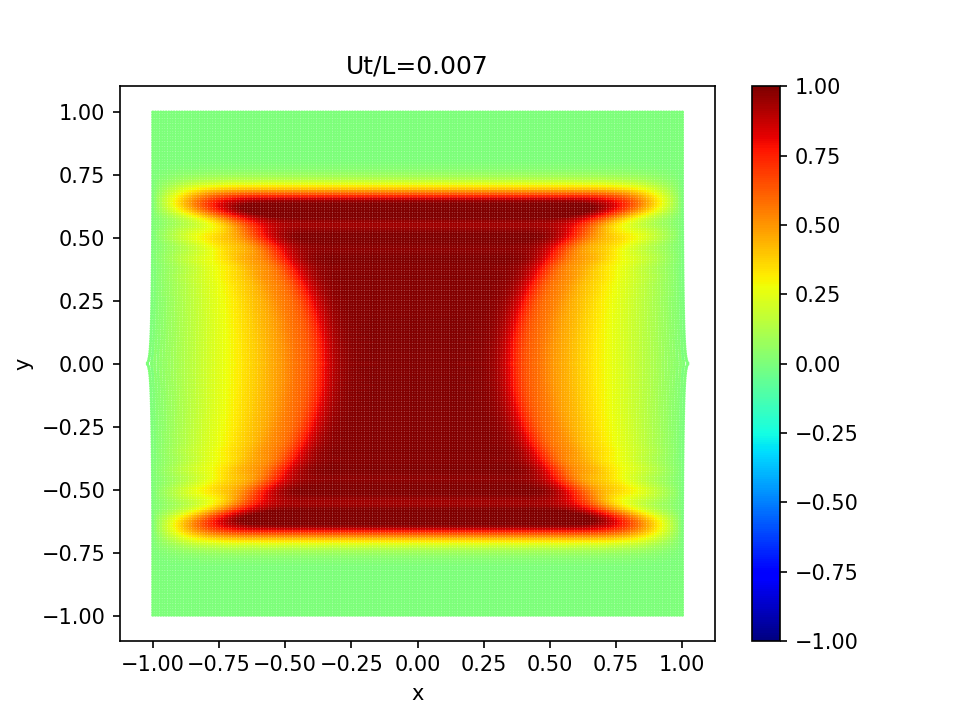}}
  \qquad
  \begin{minipage}{6cm}
    \includegraphics[width=8cm]{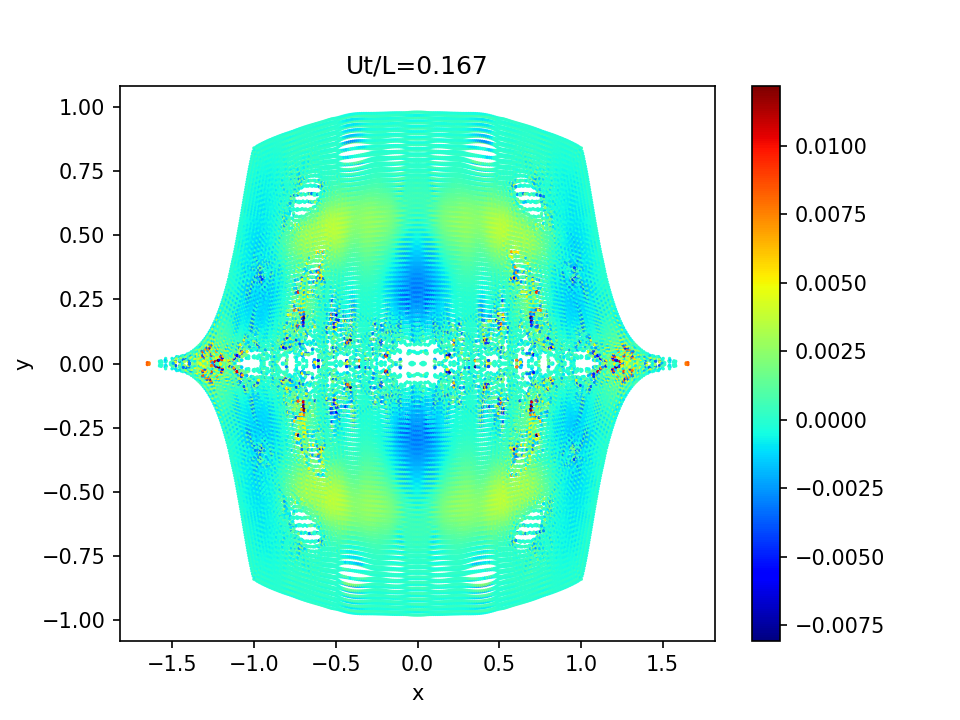}
  \end{minipage}
  \caption{Particle distribution and pressure ($p/\rho c_s U$) at $Ut/L=0.007$
    (left) and $ Ut/L = 0.167$ (right) for simulation with the standard EDAC
    scheme without any artificial viscosity.}
\label{fig:two-blocks:edac-noavisc}
\end{figure}

\begin{figure}
  \centering
  \parbox{6cm}{\includegraphics[width=8cm]{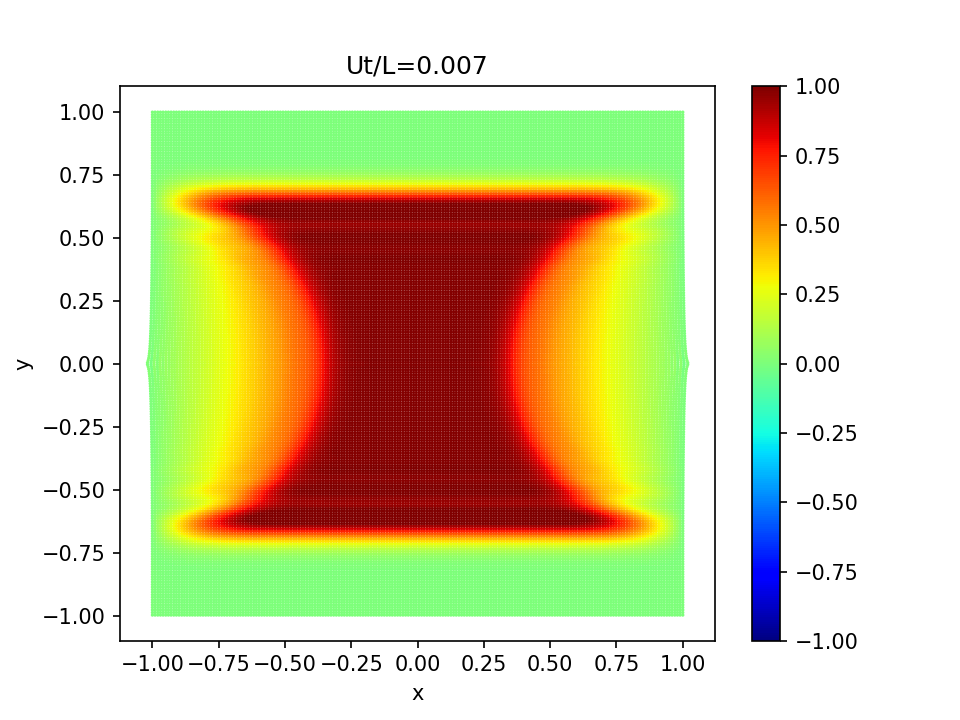}}
  \qquad
  \begin{minipage}{6cm}
    \includegraphics[width=8cm]{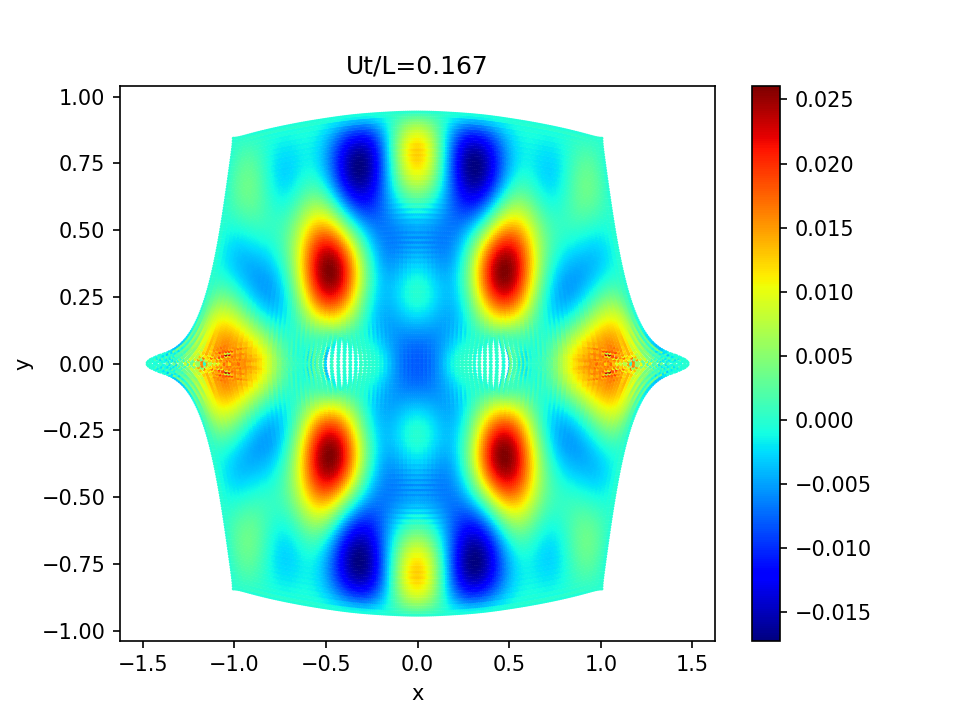}
  \end{minipage}
  \caption{Particle distribution and pressure at $Ut/L=0.007$ (left) and
    $Ut/L=0.167$ (right) for simulation with the standard EDAC scheme with
    artificial viscosity coefficient $\alpha =0.1$.}
\label{fig:two-blocks:edac}
\end{figure}

\begin{figure}
  \centering
  \parbox{6cm}{\includegraphics[width=8cm]{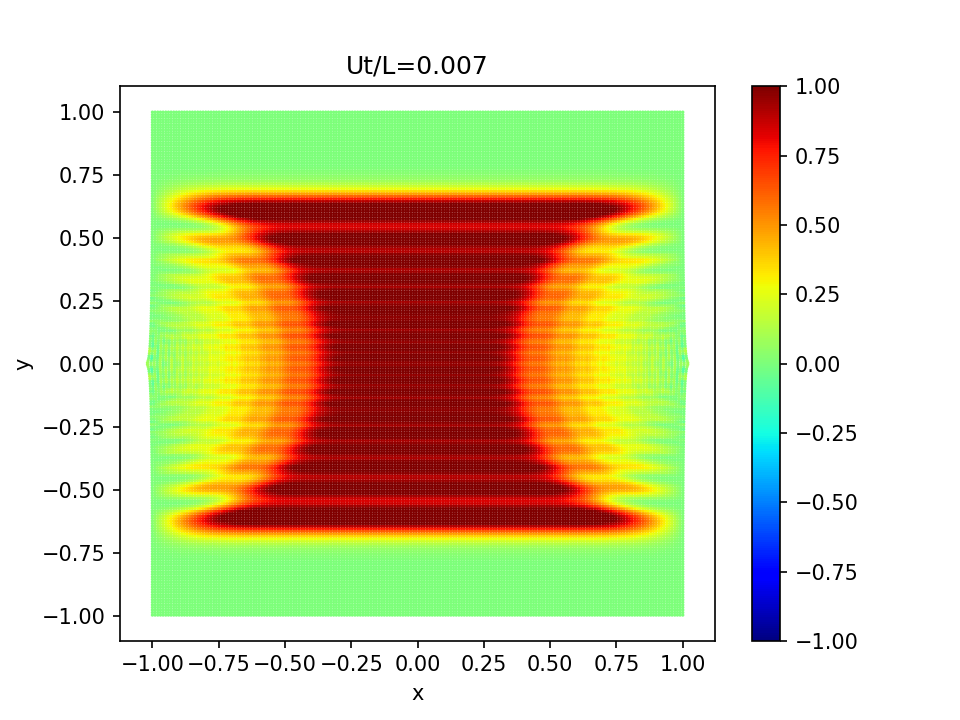}}
  \qquad
  \begin{minipage}{6cm}
    \includegraphics[width=8cm]{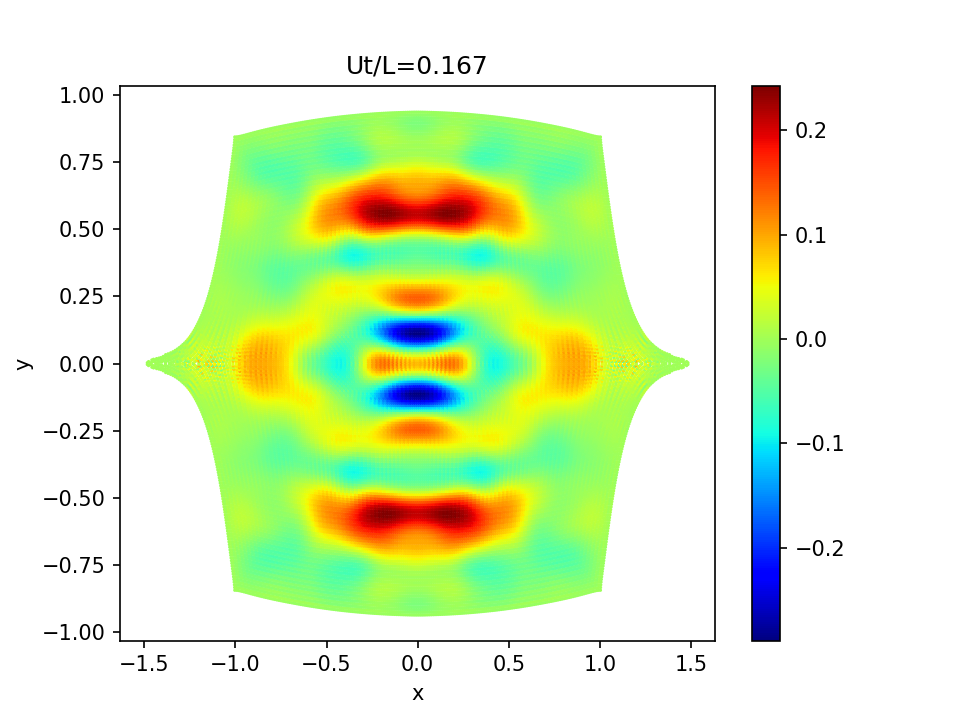}
  \end{minipage}
  \caption{Particle distribution and pressure at $Ut/L=0.007$ (left) and
    $Ut/L=0.167$ (right) for simulation with the standard WCSPH scheme with
    artificial viscosity coefficient $\alpha =0.1$.}
\label{fig:two-blocks:wcsph}
\end{figure}

The case of two rectangular blocks of water impacting is considered next. A
detailed study of this problem has been performed by
\citet{marrone:water-impact:jfs:2015} in which they use a fully compressible,
Riemann-Solver type SPH formulation and compare the results with a level-set
finite volume method. The problem involves two blocks of water, each with side
$H$ and height $L$, that are stacked vertically at $t=0$, with the interface
at $y=0$. The top block moves down with the $y$-component of velocity $v=-U$
and the bottom moves up with velocity $v=U$. There is no acceleration due to
gravity and the fluid is treated as inviscid and incompressible. Surface
tension is not modeled. The surface of the blocks is treated as a
free-surface. The problem is simulated using the standard EDAC scheme and also
the WCSPH scheme. In the present case $L=1m$, $H=2m$, $U=1m/s$ and $\rho=1.0
kg/m^3$. The Mach number is chosen to be $0.01$. For the WCSPH scheme,
$\gamma=1$. A quintic spline kernel is used for both schemes with $h=\Delta x$
and $L/\Delta x = 100$. As considered in \cite{marrone:water-impact:jfs:2015},
the normalized pressure distribution ($p/\rho c_s U$) is shown at $t^* =
Ut/L=0.007$ and at $t^* = Ut/L=0.167$. When this case is run without any
artificial viscosity, the traditional WCSPH scheme does not run successfully
until the desired time. There are large pressure oscillations.
Fig.~\ref{fig:two-blocks:wcsph-noavisc} shows the particle distribution and
pressure for the non-dimensionalized times of $t^* = 0.007$ (left) and $t^* =
0.1$ (right). In contrast, the EDAC case runs fairly well and the results are
shown in Fig.~\ref{fig:two-blocks:edac-noavisc}. Initially, the pressure is
comparable to the results in \cite{marrone:water-impact:jfs:2015}, however,
the lack of any artificial viscosity results in small pressure oscillations at
the final time and some cavitation. In Fig.~\ref{fig:two-blocks:edac} the same
case is simulated with an artificial viscosity with $\alpha=0.1$. This
produces fairly good results. It is easy to see that in all cases, the new
scheme produces much less pressure oscillations. It is worth noting that while
the WCSPH scheme requires the use of artificial viscosity for the simulation
to complete, it displays high-frequency pressure oscillations as can be seen
in Fig.~\ref{fig:two-blocks:wcsph}, where the artificial viscosity parameter
$\alpha = 0.1$ was used for the WCSPH scheme. These results clearly show the
superiority of the new scheme. In terms of performance, the WCSPH simulation
takes about 211 seconds whereas the EDAC takes about 303 seconds. This is
primarily because the EDAC implementation has not been fully optimized and due
to the additional summation density computation that the EDAC requires.

\subsection{Dam-break in two-dimensions}
\label{sec:db}

The two-dimensional dam break over a dry bed is considered next. Results are
instead compared with a standard SPH implementation. The suggested corrections
of \citet{hughes-graham:compare-wcsph:jhr:2010} and
\citet{marrone-deltasph:cmame:2011} are also employed in the implementation of
the standard SPH scheme as provided in PySPH. In the current work, only the
corrections of \citet{hughes-graham:compare-wcsph:jhr:2010} are used. The
delta-SPH corrections of \citet{marrone-deltasph:cmame:2011} do not affect the
present results.

The problem considered is as described in \citet{wcsph-state-of-the-art-2010}
with a block of water $1m$ wide and $2m$ high, placed in a vessel of length $4
m$. The block is released under the influence of gravity which is assumed to
be $-9.81m/s^2$. The particles are arranged as per a
staggered grid as is suggested for the standard SPH formulation by
\citet{wcsph-state-of-the-art-2010} with $h=0.0156$. Artificial viscosity is
used for the WCSPH implementation with a value of $\alpha=0.1, \beta=0.0$. The
standard Wendland quintic kernel is used for WCSPH case with $h = 1.3 \Delta
x$.

For the EDAC implementation, a uniform regular distribution of particles is
used as done in \cite{Adami2012}. No artificial viscosity or XSPH correction
is employed. A quintic spline kernel is used with $h=\Delta x$. The value of
$\alpha$ for the EDAC equation is set to $0.5$. The only change to the
implementation is a clamping of the boundary pressure to non-negative values
so as to prevent the fluid from sticking to the walls. At the highest
resolution the EDAC simulation uses 8192 fluid particles whereas the WCSPH use
27889 particles due to the staggered grid arrangement.

\begin{figure}[htpb]
\centering
\includegraphics[width=10cm]{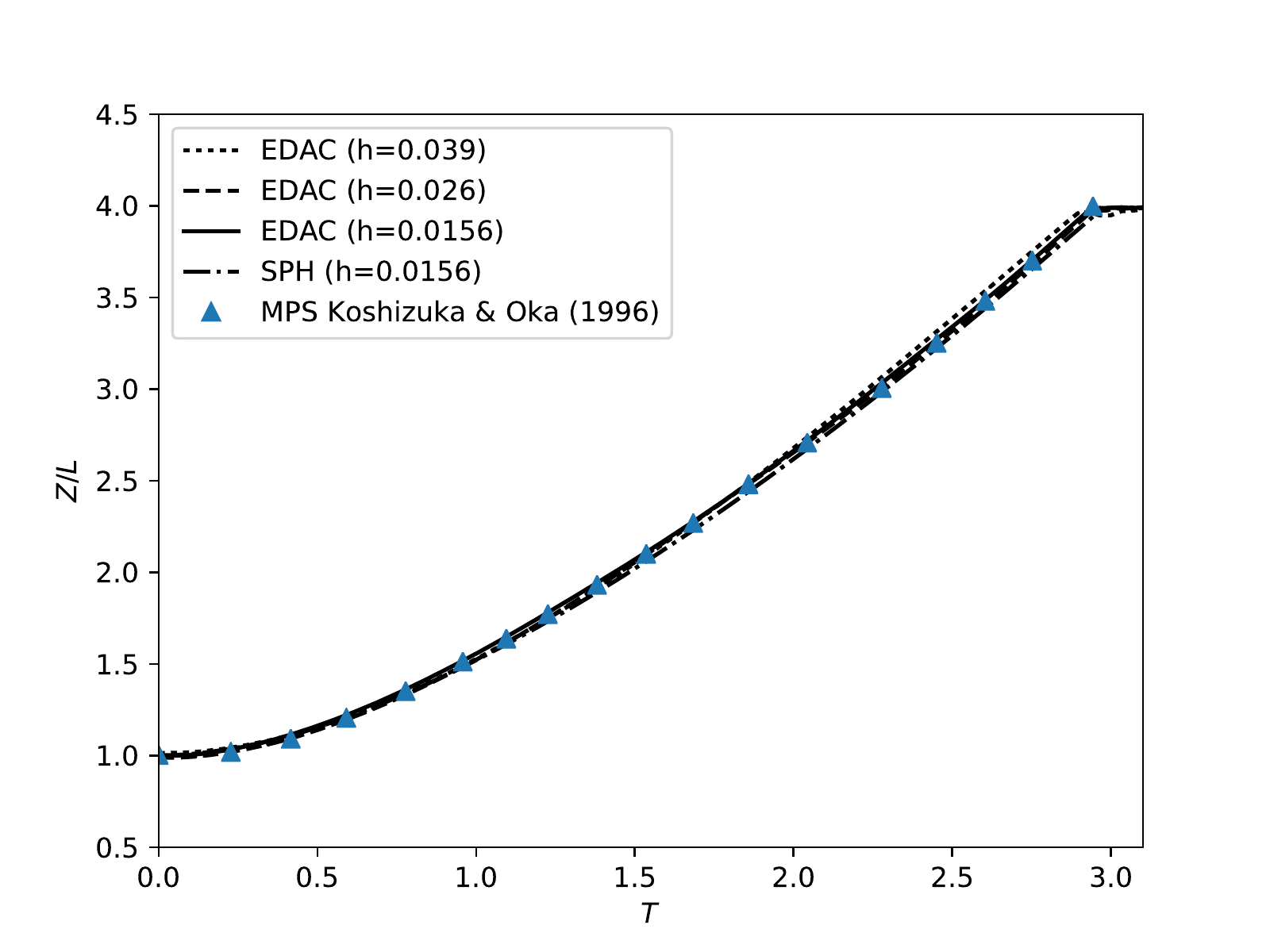}
\caption{Position of the toe of the dam as a function of time compared with
  the MPS simulation of \cite{koshizuka_oka_mps:nse:1996}. $Z$ is the distance
  of the toe of the dam from the left wall and $L$ is the initial width of the
  dam.}
\label{fig:db:toe}
\end{figure}

To compare the results, the position of the toe of the dam versus time is
plotted and compared with the results of the Moving Point Semi-implicit (MPS)
scheme of~\cite{koshizuka_oka_mps:nse:1996}. The results are plotted in
Fig.~\ref{fig:db:toe}. As can be seen, the results of the new scheme compare
well with the MPS results and the WCSPH formulation. The agreement is very
good.

\begin{figure}[htpb]
    \centering
    \begin{subfigure}[b]{0.48\textwidth}
      \includegraphics[width=1.1\textwidth]{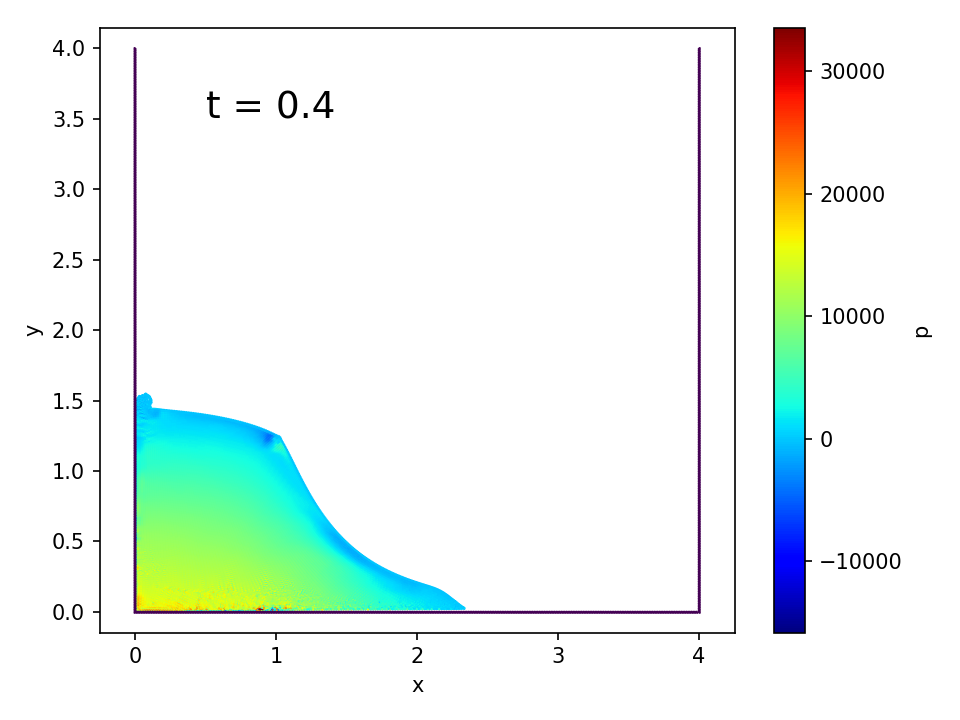}
    \end{subfigure}
    ~
    \begin{subfigure}[b]{0.48\textwidth}
      \includegraphics[width=1.1\textwidth]{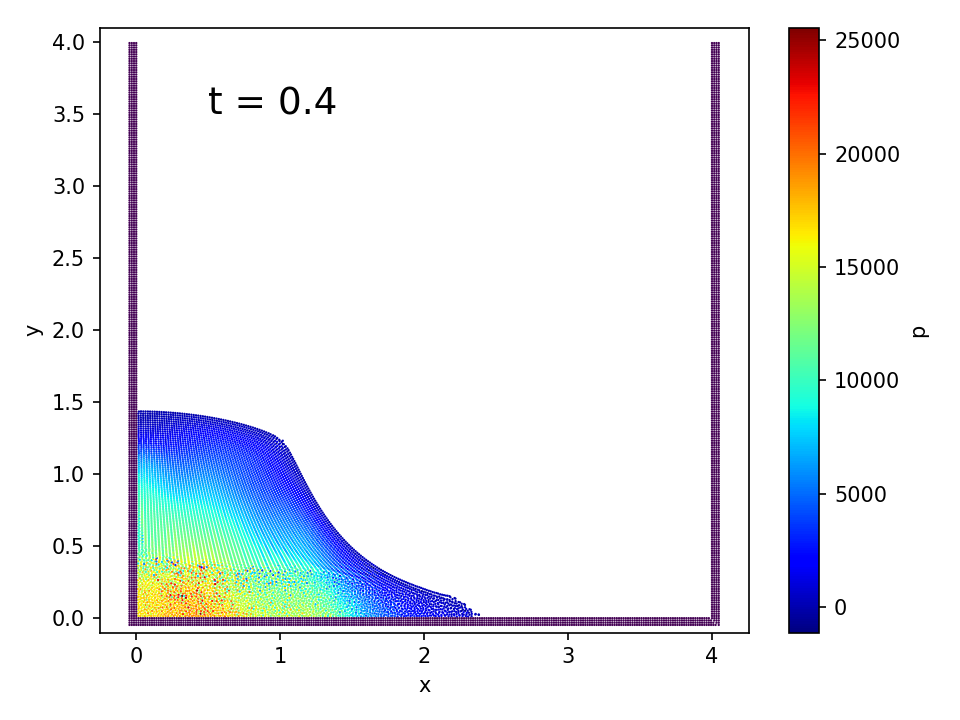}
    \end{subfigure}

    \begin{subfigure}[b]{0.48\textwidth}
      \includegraphics[width=1.1\textwidth]{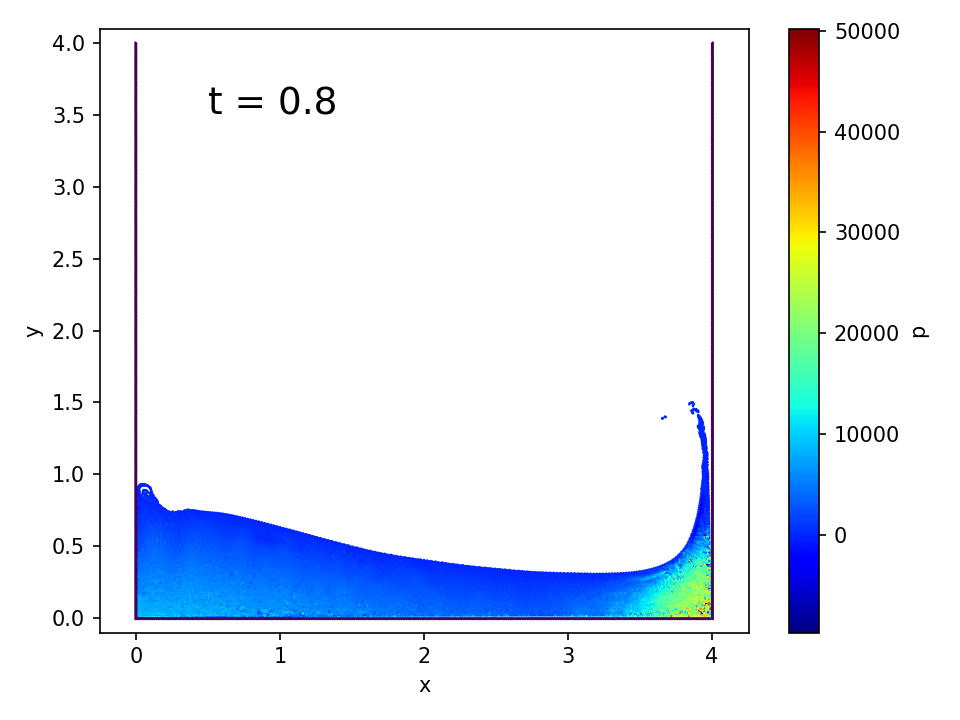}
      \caption{WCSPH}
    \end{subfigure}
    ~
    \begin{subfigure}[b]{0.48\textwidth}
      \includegraphics[width=1.1\textwidth]{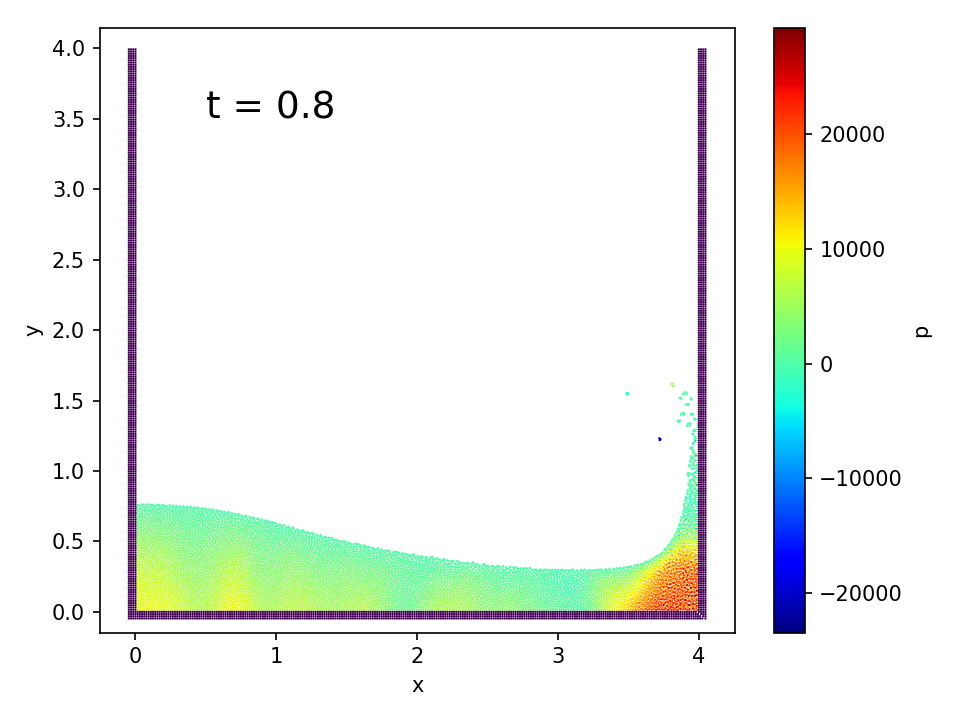}
      \caption{EDAC}
    \end{subfigure}

    \caption{Particle distribution for the two-dimensional dam break problem, for the
      WCSPH and the EDAC cases.  The WCSPH cases are on the left and the EDAC
      on the right.  The color indicates the pressure.  The top row is at a
      time of 0.4 seconds and the bottom at 0.8 seconds. }
    \label{fig:dam_break_p}
\end{figure}

Fig.~\ref{fig:dam_break_p} shows the distribution of particles with the color
indicating the pressure. The left panel shows the results obtained using the
WCSPH scheme and the right shows that of the EDAC scheme. The top row is at
0.4 seconds and the bottom at 0.8 seconds. The fluid near the left wall is
better behaved in the EDAC case and the surface is smooth. The EDAC scheme
displays a larger amount of splashing due to the lack of any artificial
viscosity in the momentum equation.  Both schemes appear to show some noise
near the left bottom wall at $t=0.4s$.  The pressure magnitudes in the WCSPH
case are much larger than those of the EDAC scheme.

\begin{figure}[htpb]
    \centering
    \begin{subfigure}[b]{0.48\textwidth}
      \includegraphics[width=1.1\textwidth]{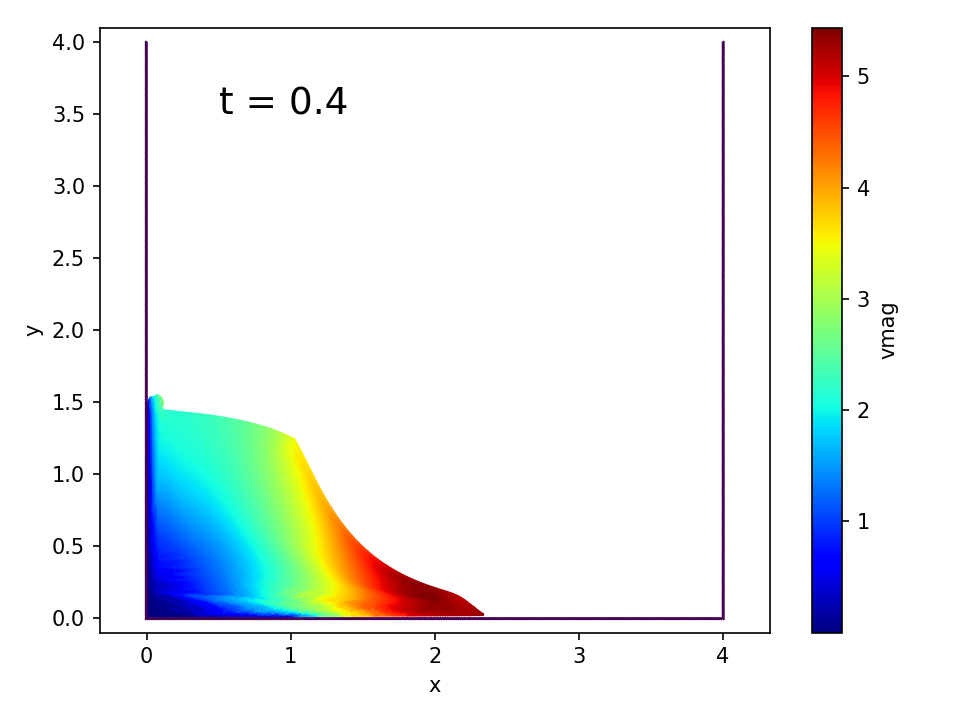}
    \end{subfigure}
    ~
    \begin{subfigure}[b]{0.48\textwidth}
      \includegraphics[width=1.1\textwidth]{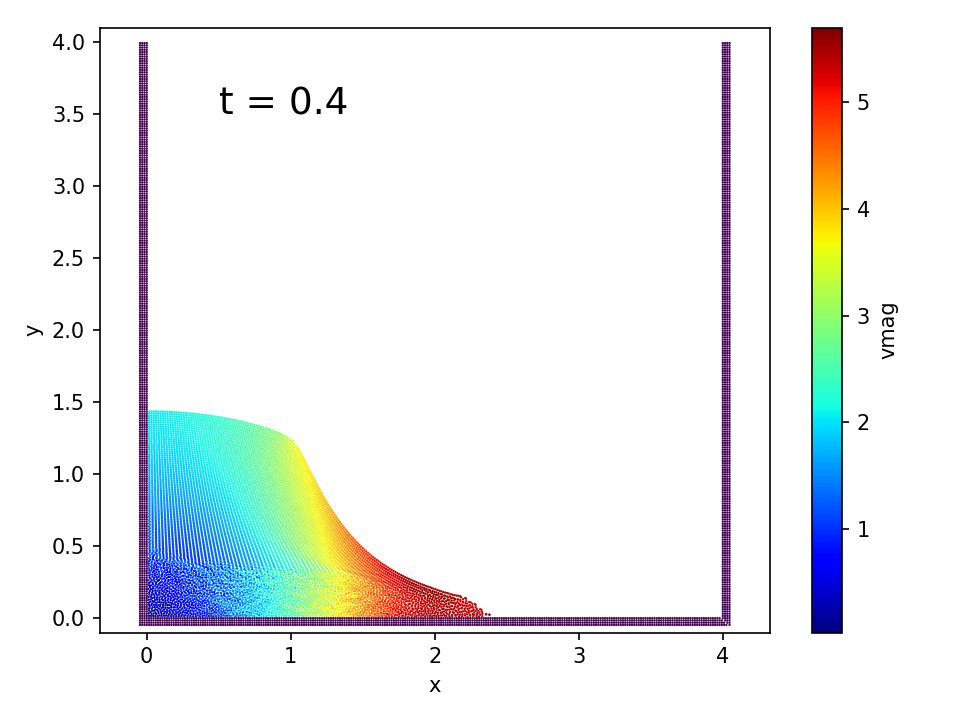}
    \end{subfigure}

    \begin{subfigure}[b]{0.48\textwidth}
      \includegraphics[width=1.1\textwidth]{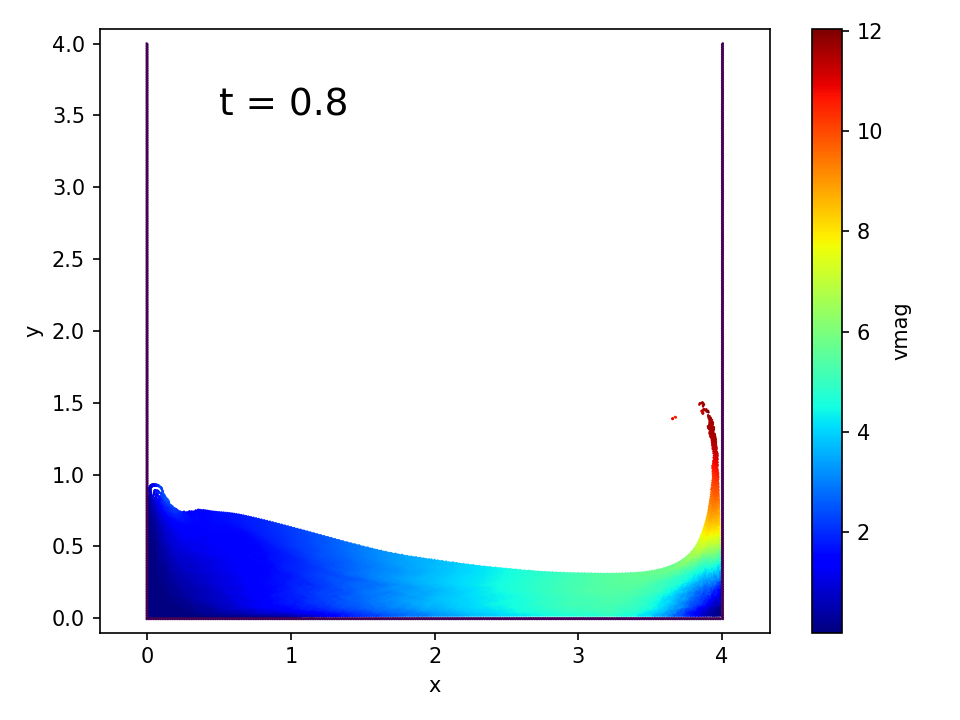}
      \caption{WCSPH}
    \end{subfigure}
    ~
    \begin{subfigure}[b]{0.48\textwidth}
      \includegraphics[width=1.1\textwidth]{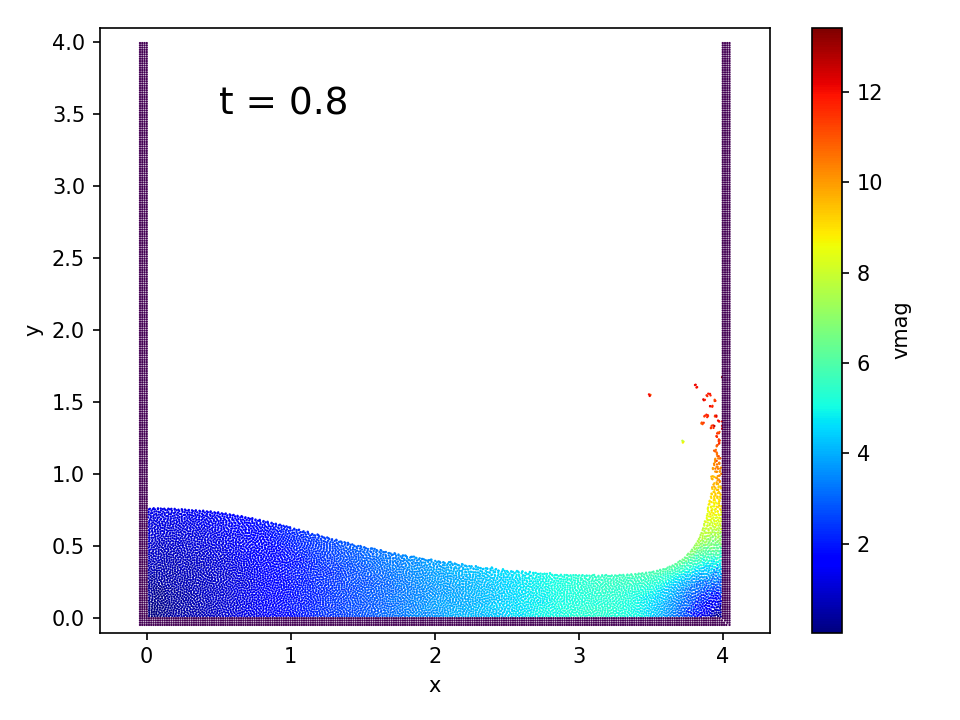}
      \caption{EDAC}
    \end{subfigure}

    \caption{Particle distribution showing the velocity magnitude for the
      two-dimensional dam break problem, for the WCSPH and the EDAC cases. The
      WCSPH cases are on the left and the EDAC on the right. The top row is at
      a time of 0.4 seconds and the bottom at 0.8 seconds. }
    \label{fig:dam_break_vmag}
\end{figure}

Fig.~\ref{fig:dam_break_vmag} shows the velocity magnitude of the particles.
The results are fairly similar.  These results show that the new scheme works
well for this problem.  The EDAC scheme does not require the use of artificial
viscosity in the momentum equation or the use of the XSPH correction.

The benchmarks above show that the new scheme produces good results for
internal and external flow problems.

\section{Conclusions}
\label{sec:conclusions}

In this work, the Entropically Damped Artificial Compressibility scheme of
\citet{Clausen2013} is applied to SPH. Two flavors of the new scheme are
developed, one called the EDAC TVF scheme which is suitable for internal
flows, and the other called the standard EDAC scheme which is suitable for
external flows. The key elements of the EDAC TVF scheme are the use of the
EDAC equation to evolve the pressure, the use of the transport velocity
formulation of \citet{Adami2013}, and, importantly, a pressure correction as
suggested by \citet{sph:basa-etal-2009}. This scheme outperforms the TVF
scheme for the Taylor Green vortex problem at various Reynolds numbers. The
scheme performs very well for a variety of other internal flow problems. The
standard EDAC scheme is easy to apply to external flow problems and to
free-surface flows. The method produces results that are better than the
standard SPH. The pressure distribution is smoother and more accurate. It does
not require the use of artificial viscosity and is relatively simple to
implement. It is seen that a judicious choice of the viscosity for the
pressure equation is important. A heuristic expression is suggested that
appears to work well for all the simulated problems. While this viscosity
introduces a new parameter, our computations suggest that this parameter does
not need to be tuned for different problems and a value of 0.5 or 1.0 works
well for a variety of Reynolds numbers and problems. A fully working
implementation of the scheme and all the benchmarks in this paper are made
freely available in order to encourage reproducible computational science.

\section*{Acknowledgments}

The authors are grateful to the anonymous reviewers for their comments that
have made this manuscript better.




\section*{References}
\bibliographystyle{model6-num-names}
\bibliography{references}


\end{document}